\DocumentMetadata{%
	pdfstandard=A-3u,
	pdfversion=1.7,
	lang=en-US,
}

\documentclass[nocopyright,upint,varvw,barcolor=Goldenrod3,mathalfa=cal=euler,balance,hyphenate,arxiv,nolists]{asmejour}

\def\papertitle{Surrogate-Based Co-Design Coupling Analysis for Floating Offshore Wind Turbines}
\def\paperkeywords{design coupling analysis (DCA), control co-design (CCD), floating offshore wind turbines (FOWT), surrogate model-based optimization (SMBO), sensitivity}

\hypersetup{%
	pdfauthor={Yong Hoon Lee},
	pdftitle={\papertitle},
	pdfkeywords={\paperkeywords},
	pdfsubject = {Surrogate-Based Co-Design Coupling Analysis for Floating Offshore Wind Turbines},
}

\JourName{Mechanical Design}
\PaperYear{}

\begin{document}

\SetAuthorBlock{Elena Fern{\'a}ndez~Bravo\CoFirstAuthor}{Department of Aerospace Engineering,\\
   University of Illinois at Urbana-Champaign,\\
   104 South Mathews Avenue,\\
   Urbana, IL 61801, USA \\
   email: elenaf3@illinois.edu}

\SetAuthorBlock{Sunil Tamang\CoFirstAuthor}{Department of Mechanical Engineering,\\
   The University of Memphis,\\
   312 Engineering Science Building,\\
   Memphis, TN 38152, USA \\
   email: stamang1@memphis.edu}

\SetAuthorBlock{Yong Hoon Lee\CorrespondingAuthor}{Department of Mechanical Engineering,\\
   The University of Memphis,\\
   312 Engineering Science Building,\\
   Memphis, TN 38152, USA \\
   email: yhlee@memphis.edu}

\SetAuthorBlock{James T. Allison}{Department of Industrial and Enterprise Systems Engineering,\\
   University of Illinois at Urbana-Champaign,\\
   104 South Mathews Avenue,\\
   Urbana, IL 61801, USA \\
   email: jtalliso@illinois.edu}

\title{\papertitle}

\begin{abstract}
This work presents a design coupling analysis (DCA) framework to investigate the interactions among control and plant design variables in floating offshore wind turbine (FOWT) and to support the formulation of tractable control co-design (CCD) optimization strategies. DCA provides quantitative information that reveals the relationships and dependencies among design variables and to objective function, enabling improved design variable selection, identification of dominant variables that drive system interactions, and informed selection of optimization solution strategies. However, applying DCA to complex systems is challenging because the models used to describe their dynamics are computationally expensive, and constructing DCA information requires exhaustive model evaluations and optimizations. Here, a surrogate model of the FOWT system is employed to make the repeated model evaluations required for DCA computationally feasible. Using this framework, the bidirectional couplings between control and plant design variables, as well as the couplings among plant design variables, are estimated. The results reveal strong interactions among various design variables, and identify the most influential plant design variables affecting system performance. These insights guide the development of two DCA-based optimization strategies for large CCD problems: a sequential decomposition approach that preserves dominant design variable couplings while reducing problem size at each stage, and a reduced dimensional optimization approach that focuses on collectively the most influential variables. The results demonstrate that these strategies significantly reduce computational complexity while achieving solutions comparable to those obtained through full simultaneous optimization, demonstrating the value of DCA for understanding and solving complex design problems.

\end{abstract}

\date{}
\maketitle

\section{Introduction}

Design coupling quantifies how a change in one design decision alters the optimal values of other design decisions, going beyond traditional sensitivity analysis that merely quantifies model responses. Analysis on design coupling provides profound knowledge on how we should make design decisions considering quantifiable interactions among design variables and their optimal solutions. Similarly to sensitivity analysis information, the design coupling information can be represented as a Jacobian matrix of design variables; however, it represents information that is fundamentally distinct to the Jacobian matrix of partial derivatives of model responses with respect to design variables~\cite{Chinthoju2024eVTOL}.

Unlike analysis coupling or sensitivity analysis, design coupling analysis (DCA) provides how the change of a variable impacts changes in not only responses of other variables but also optimal decisions on these variables. Chinthoju~\textit{et al.}~\cite{Chinthoju2024eVTOL} first introduced the design coupling information in multidisciplinary design optimization (MDO) as a Jacobian matrix and visualized how decisions on an arbitrary design variable impacts optimal decision in the rest of design variables. The study found that the Jacobian is not always symmetric, implying that the decision impacts are not always bidirectionally symmetric. Fern{\'a}ndez~Bravo and Allison~\cite{FernandezBravo2024SatelliteMDO} expanded their work by also considering the sensitivity of the objective function in the DCA and by providing different approaches to estimate the Jacobian matrices.

In offshore wind energy, control co-design (CCD) has emerged as a promising direction for technical innovation, enabling simultaneous optimization of design variables across plant and control disciplines~\cite{Pao2024AnnuRev}. Notably, recent studies on CCD have shown substantial leaps in both technical and economical performance by accounting for interdisciplinary couplings, highlighting the potential of this approach to significantly enhance the efficiency and cost-effectiveness of offshore wind farms in practical industry applications~\cite{Lee2025FloatVAWT, Lee2025AppliedEnergy, Bayat2025OE, sundarrajanOpenLoopControlCoDesign2024, ABBAS2024ccd}. Nested CCD formulations achieved efficient plant and control co-design by nesting faster dynamic computation inside the plant design formulation~\cite{Cui2021, sundarrajanOpenLoopControlCoDesign2024, Bayat2025OE}. Exploring a wide range of control design space in early-stage design problems achieves large performance improvements over existing designs~\cite{Lee2025FloatVAWT, Sakib2024}. Challenging structural constraints could better be addressed by CCD than control design domain alone~\cite{Deshmukh2016}. To address CCD problems associated with the stochastic and non-uniform nature of wind energy, various modeling and computational methods, including deep learning (DL), reliability-based MDO (RBMDO), sequential optimization and reliability assessment (SORA), and extended random search, have been investigated, contributing to improved design understanding in wind energy system~\cite{MengKrigingAssistedRBMDO2023, ChenReviewOffshoreWind2022, ChoeSequenceBasedModeling2021, Cui2021, FengDesignOptimizationOffshore2017, MuskulusDesignOptimizationSupport2014}.

Design coupling in CCD problems can be estimated in several different ways. Fathy~\textit{et al.}~\cite{fathyCouplingPlantController2001, fathyCombinedPlantControl2003} obtained the coupling term by taking the difference between the CCD problem's optimality and separated problems' optimality conditions. For the separated problems, they used Karush-Kuhn-Tucker (KKT) conditions for plant optimality and Pontryagin conditions for control optimality. This coupling measure has also been used in Refs.~\cite{shabdeOptimumControllerDesign2008, ulsoySmartProductDesign2019}. The same approach is taken in Ref.~\cite{fathyCombinedPlantControl2004}, where the authors provide a coupling measure under the assumption that the system dynamics are linear and time-invariant. Peters, Papalambros, and Ulsoy~\cite{petersRelationshipCouplingControllability2015} showed that, for some problem formulations, the coupling strength can be determined a priori using the controllability Grammian. This method has the advantage of not requiring the solution of the system of KKT conditions. One key assumption in this earlier work is that the coupling is unidirectional, neglecting the effect of the control problem on the plant design problem. This previous work was extended to account for bidirectional coupling for specific cases~\cite{petersMeasuresCouplingArtifact2009}, however, the control design variables considered throughout the paper were scalar quantities. Fern{\'a}ndez~Bravo, Ornik, and Allison~\cite{fernandezbravoNumericalEstimationBidirectional2024} estimated the bidirectional control and plant design variable coupling strength, including when the control inputs are time-varying, by calculating the optimal plant and optimal control sensitivities. These sensitivities were used to guide problem formulation decisions and identify the most impactful plant and controller variables.

A major challenge in obtaining design coupling information prior to formulating and solving CCD problems lies in the high demand in computational resource that is required to calculate the partial derivatives of the optimal solution with respect to perturbations in each design variable of interest. While solving a massive number of sub-optimization problems to reduce the full-scale design optimization problem size seems impractical, the use of efficient surrogates of full-fidelity model's responses practically enable this approach. These surrogates, which provide higher computational efficiency, can still
yield valuable insights for analyzing design coupling information. Here, surrogate-based optimization plays a critical role in delivering efficient yet informative coupling insights for CCD applications.

Surrogate models have long been employed to mitigate the computational expenses of solving CCD problems. Many studies adopt a black-box approach, replacing simulation models and performance metrics with computationally efficient surrogates to approximate system behaviors~\cite{Wang2020SBCCD, qiaoNewSequentialSampling2021, zhangNovelSurrogateModelBased2022}. More advanced surrogate-based optimization methods specifically tailored to CCD problems have also emerged, such as the derivative function surrogate modeling (DFSM) technique, which captures dynamic state derivatives~\cite{Lee2025AppliedEnergy, sundarrajanOpenLoopControlCoDesign2024, sundarrajanUsingHighFidelityTimeDomain2023, Deshmukh2017DFSM}, and approaches that discretize continuous-time control parameters with surrogate modeling process and reconstructing full time histories with it~\cite{Wang2020SBCCD}. Data-driven strategies leveraging surrogates to create and iteratively refine design optimization formulations (such as defining constraint boundaries) are also introduced in the literature~\cite{Lee2019SMO, Malak2010SVDD}. Additionally, many wind turbine CCD research articles have shown the effectiveness of employing surrogate-based optimization~\cite{Lee2025FloatVAWT, LucasFrutuoso2025AdvWindFarm, sundarrajanUsingHighFidelityTimeDomain2023, zhangNovelSurrogateModelBased2022, qiaoNewSequentialSampling2021, Deshmukh2017DFSM}.

This study employs surrogate-based DCA to efficiently provide compact design formulations for floating offshore wind turbines (FOWTs). These systems involve several coupled physical analysis disciplines, including aeroelasticity, multi-body structural dynamics, hydrodynamics, and controls~\cite{Bayat2026wtccdapen}. The system is modeled with the Wind Energy with Integrated Servo-control (WEIS), an open-source project that is currently being developed by the team including authors, primarily led by National Laboratory of the Rockies (NLR)~\cite{jonkmanFunctionalRequirementsWEIS2021}. The number and complexity of the disciplines make evaluating the model very computationally expensive, and given the high number of model evaluations required for the DCA, this task becomes infeasible.  A surrogate model is developed and used to perform the DCA.

The article is structured as follows: Section~\ref{sec:fowt_model} provides an overview of the FOWT modeling framework. Sections~\ref{sec:sbo} and~\ref{sec:DC_est} detail the numerical estimation of design coupling supported by surrogate-based optimization. The described methodology is applied to an example in Sec.~\ref{sec:results}. Finally, Sec.~\ref{sec:conclusions} provides concluding remarks and lays out future work opportunities. 

\section{Floating Offshore Wind Turbine Modeling}
\label{sec:fowt_model}

This section provides an overview of the modeling framework used to represent the FOWT design explored and analyzed in this study.

\subsection{Floating Offshore Wind Turbine Dynamics}

Modeling the dynamics of FOWTs is inherently multidisciplinary, involving strongly coupled aero-hydro-elasto-servo interactions among the rotor blades, tower, floating platform, mooring lines, control system, and surrounding environmental conditions~\cite{Matha2016FOWT}. Turbulent inflow wind with gust and directional variations, incident waves, and ocean currents influence aerodynamic loading on the rotor, hydrodynamic forces acting on the floating platform, and forces transmitted through the mooring system and fairleads, leading to strongly coupled system responses across multiple physical domains. These interactions are further influenced by the control system, which actively regulates wind turbine dynamics in response to both system states and environmental excitations. Over the past decades, modeling approaches for FOWTs have evolved to reflect these complex couplings with increasing physical fidelity while balancing computational efficiency and practical applicability. As a result, a broad spectrum of modeling tools has emerged, ranging from high-fidelity simulations for detailed physics to low-fidelity approximations designed for rapid analysis, each offering tailored trade-offs between accuracy and computational demand.

High-fidelity methods, such as fluid-structure interaction (FSI) incorporated into computational fluid dynamics (CFD) simulations, are typically used for fundamental scientific research due to their prohibitive modeling and computational costs. Several studies embed high-fidelity methods with relatively lower-fidelity simulations, mitigating the computational cost for accurate simulation of severe operating conditions, such as extreme gusts and storm-level metocean states~\cite{Johlas2019LES, CampanaAlonso2023OF2}. Mid-fidelity computational models, such as OpenFAST~\cite{Jonkman2013FAST}, QBlade~\cite{Marten2020QBlade}, and Bladed~\cite{Beardsell2018Bladed}, are more frequently used both in fundamental scientific research and practical development and design projects because of their accessibility in terms of computational cost and profound information that these models can provide. Lower-fidelity or reduced-order models are also widely adopted in scenarios requiring computational efficiency and simplicity, such as early-stage rapid design exploration, real-time digital twin deployment, and control-oriented studies. Widely adopted methods include linear or linearized time-domain models~\cite{JonkmanFullSystemOpenFAST2018, sundarrajanOpenLoopControlCoDesign2024} and frequency-domain models~\cite{hallOpenSourceFrequencyDomainModel2022}. Frequency-domain models are particularly attractive for early sizing and load-case screening, yet only a limited number of tools are available~\cite{OtterReviewFOWT2021}.

This study employs mid- and low-fidelity modeling approaches for FOWTs. To extract design coupling information efficiently without excessive computational cost, surrogate models are constructed using low-fidelity frequency-domain responses from the Response Amplitudes of Floating Turbines (RAFT)~\cite{hallOpenSourceFrequencyDomainModel2022} software included in the WEIS toolset. The final design solution, derived from the formulation generated by the DCA tool, is then computed using the mid-fidelity time-domain model, OpenFAST, also embedded within the WEIS toolset. The entire process of DCA and problem formulation described in this study has been integrated into the WEIS toolset as part of its functionality.

\subsection{Baseline Wind Turbine Design} \label{sec:baselinewtdesign}

The IEA Wind 15-megawatt offshore reference wind turbine (hereinafter referred to as the IEA 15~MW), featuring a rotor diameter of 240~m and a hub height of 150~m~\cite{Gaertner2020IEA15MWrefturb}, supported on the UMaine VolturnUS-S reference semi-submersible platform (hereinafter referred to as the platform)~\cite{Allen2020UMaineSemiRef}, is used in this study as the baseline configuration for design exploration and coupling analysis.

For numerical hydrodynamic modeling simplicity, the lower pontoons that connect the outer columns to the central column are represented as circular cylinders with equivalent displaced volume, rather than the rectangular prisms specified in the original platform design definition. The reference documentation reports a total platform mass of 17,839 tonnes, including fixed and fluid ballast~\cite{Allen2020UMaineSemiRef}. However, the RAFT input model used in this work has a reduced mass of 15,093 tonnes. This discrepancy arises from modeling simplifications and limitations of the frequency-domain approach, which cannot capture all detailed structural features of the full design. The DCA presented in this work uses the baseline RAFT model as the starting configuration.

\subsection{Wind Energy with Integrated Servo-control ({WEIS})} \label{sec:weistoolset}

The WEIS toolset employed in this work is an MDO framework that integrates multiple wind turbine analysis and design software packages, enabling flexible modeling of general wind turbine configurations and the solution of CCD problems for both onshore and offshore systems~\cite{jonkmanFunctionalRequirementsWEIS2021}. The primary design analysis and optimization routine is based on WISDEM, a systems engineering software for wind turbines, but expanded to orchestrate with multiple fidelity-level simulation codes, such as OpenFAST, TurbSim (a stochastic, full-field, turbulent-wind simulator), ROSCO (an open, modular, and fully adaptable wind turbine controller), HAMS (a boundary element method for hydrodynamic analysis of submerged structures~\cite{Liu2019HAMS}), and RAFT, and CCD techniques, such as DTQP (a linear-quadratic dynamic optimization solver using direct transcription and quadratic programming~\cite{herberAdvancesCombinedArchitecture, Herber2017DTQPproject}), multi-fidelity optimization \cite{Jasa2022WEISmultifidelity}, and DCA capabilities~\cite{Chinthoju2024eVTOL, fernandezbravoNumericalEstimationBidirectional2024}. The developments and findings presented in this study will be incorporated back into the ongoing advancement of the WEIS toolset.

\subsection{Frequency-Based Dynamic Response Model}

RAFT is an open-source frequency-domain dynamic response model developed to efficiently compute the steady-state behavior of FOWTs~\cite{hallFrequencyDomainModelingFloating2025}. It employs a linearized formulation to solve for mean platform offsets and dynamic responses in frequency-dependent harmonic components, $\bm{\hat{\xi}}\left(\omega\right)$, with respect to a mean equilibrium state, $\bm{\bar{\xi}}$, under combined wind and wave loading, considering the six rigid-body degrees of freedom (surge, sway, heave, roll, pitch, and yaw). External loads are represented analogously, decomposed into mean components, $\mathbf{\bar{f}}$, and harmonic amplitudes, $\mathbf{\hat{f}}(\omega)$.

RAFT models the floating structure as an assembly of member elements, with hydrodynamic forces evaluated using a strip theory formulation based on Morison's equation to account for added mass, damping, and wave excitation~\cite{morisonForceExertedSurface1950}. Mooring effects are incorporated through MoorPy, a quasi-static mooring solver~\cite{hallMoorPyQuasiStaticMooring2021} that uses the platform hydrostatic properties and mean environmental loads to determine equilibrium offsets, mean restoring forces, and linearized mooring stiffness contributions for the frequency-domain analysis. Rotor aerodynamics in RAFT is modeled using CCBlade, a steady-state blade element momentum solver~\cite{ningSimpleSolutionMethod2014}, that takes hub height, shaft tilt angle, hub wind speeds, rotor RPM, and blade pitch as inputs and returns rotor power, thrust, torque, and their derivatives as outputs for linearization.

\section{Surrogate-Based Optimization}
\label{sec:sbo}

DCA requires repeatedly solving constrained optimization problems under systematic perturbations of candidate design variables. To reduce the computational burden associated with these repeated evaluations and optimizations, we have constructed computationally inexpensive surrogate models of the RAFT model responses of interest. In this work, the surrogate model is used exclusively to compute design coupling information represented by Jacobian matrices of optimal design sensitivities discussed in Sect.~\ref{sec:DC_est}. Final design verification can still be evaluated with higher-fidelity tools (e.g., OpenFAST within WEIS), as described in Sec.~\ref{sec:fowt_model}.

\subsection{Surrogate Model Formulation}

Let $\mathbf{x} \in \mathbb{R}^{N_x}$ denote the plant and control design variable vector, $\mathbf{p} \in \mathbb{R}^{N_p}$ fixed parameter vector, and $\mathbf{y} \in \mathbb{R}^{N_y}$ the vector of model outputs of interest. In this study $\mathbf{y}$ includes quantities needed to evaluate objective and constraint functions and characterize system responses. The surrogate model is constructed to approximate each component of the system response vector independently ($y \in \mathbf{y}$), given as:
\begin{align}
    y = \mathcal{M}\left(\mathbf{x}; \mathbf{p}\right) \approx \tilde{\mathcal{M}}\left(\mathbf{x}; \mathbf{p}\right), \label{eq:surrogate}
\end{align}
where $\mathcal{M}$ represents the physics-based analysis model (i.e., RAFT, in this study), and $\tilde{\mathcal{M}}$ is the constructed surrogate model of $\mathcal{M}$. It is essential that the surrogate model provides smooth predictions across the design space and represents nonlinear trends in all output variables.

\subsection{Sparse Gaussian Process Regression} \label{sec:sgp}

Gaussian process regression is defined, starting from Eq.~\eqref{eq:surrogate}, given as:
\begin{align}
    y \approx \tilde{\mathcal{M}}\left(\mathbf{x}; \mathbf{p}\right) = \tilde{\mathcal{M}}_{\text{GP}}\left(\bar{y}\left(\mathbf{x}; \mathbf{p}\right), \kappa\left(\mathbf{x},\mathbf{x}'; \mathbf{p}\right)\right),
\end{align}
where $\bar{y}\left(\cdot\right)$ and $\kappa\left(\cdot\right)$ denote the prior mean and covariance (kernel) functions, respectively. Hereinafter, $\mathbf{p}$ is omitted in the formulation for simplicity. In this study, a squared exponential kernel is adopted. For two input points $\mathbf{x}_1$ and $\mathbf{x}_2$, the kernel is defined as:
\begin{align}
    \kappa\left(\mathbf{x}_1, \mathbf{x}_2\right) = \sigma_f^2 \exp \left(-\dfrac{1}{2l^2}\left|\left|\mathbf{x}_1 - \mathbf{x}_2\right|\right|_2^2\right),
\end{align}
where $\sigma_f^2$ is the signal variance and $l$ is the length scale parameter; together they govern the amplitude and smoothness of the kernel response. Standard Gaussian process regression provides smooth predictions and uncertainty estimates, but requires $\mathcal{O}(N^3)$ operations for a training set $\{\mathbf{X},\mathbf{Y}\}$ with $\mathbf{X}=\{\mathbf{x}_1,\cdots,\mathbf{x}_N\}$ and $\mathbf{Y}=\{y_1,\cdots,y_N\}$, due to inversion of the full $N \times N$ kernel matrix, which becomes prohibitive for large training sets.

To resolve this issue, a sparse Gaussian process (SGP) surrogate model based on the fully-independent training conditional (FITC) approximation is employed in this study~\cite{bauerUnderstandingProbabilisticSparse2016}. In the FITC formulation, a reduced set of inducing points, $\mathbf{Z}=\{\mathbf{z}_1,\cdots,\mathbf{z}_M\}$ with $M \ll N$, and corresponding latent function values, $\mathbf{u} = \hat{y}\left(\mathbf{Z}\right)$, are introduced, where $\hat{y}\left(\cdot\right)$ denotes the latent Gaussian process, and the observed outputs satisfy $y = \hat{y}\left(\mathbf{x}\right) + \varepsilon$, with $\varepsilon \sim \mathcal{N}\left(0, \sigma^2\right)$. Here, $\sigma^2$ denotes the Gaussian observation noise variance. The FITC approximation to the covariance of the observed training outputs is:
\begin{align}
&\mathbf{K}_{\text{FITC}} \equiv \mathbf{Q} + \bm{\Lambda} + \sigma^2 \mathbf{I}, \text{ with} \\
&\mathbf{Q} \equiv \mathbf{K}_{yu}\mathbf{K}_{uu}^{-1}\mathbf{K}_{uy}, \nonumber \\
&\bm{\Lambda} \equiv \text{diag}\left(\mathbf{K}_{yy} - \mathbf{Q}\right), \nonumber
\end{align}
where $\mathbf{K}_{yu}=\kappa(\mathbf{X},\mathbf{Z})$, $\mathbf{K}_{uu}=\kappa(\mathbf{Z},\mathbf{Z})$, and $\mathbf{K}_{uy}=\kappa(\mathbf{Z},\mathbf{X})$. All covariance matrices are defined with respect to the latent Gaussian process $\hat{y}\left(\cdot\right)$. Then, the FITC cross-covariance approximation is given as:
\begin{align}
    &\mathbf{q}_{*y} = \mathbf{q}_{y*}^{\textsf{T}} = \mathbf{k}_{*u} \mathbf{K}_{uu}^{-1} \mathbf{K}_{uy}, \text{ with} \\
    &\mathbf{k}_{*u} = \kappa\left(\mathbf{x}_*, \mathbf{Z}\right), \quad \mathbf{K}_{u*} = \kappa\left(\mathbf{Z}, \mathbf{x}_*\right) = \mathbf{k}_{*u}^{\textsf{T}}, \nonumber
\end{align}
then the FITC predictive distributions for the latent function $\hat{y}_*$ at a test point $\mathbf{x}_*$ are:
\begin{align}
    &\mu_{\text{FITC}*} = \bar{y}\left(\mathbf{x}_*\right) + \mathbf{q}_{*y} \mathbf{K}_{\text{FITC}}^{-1}\left(\mathbf{Y} - \bar{y}\left(\mathbf{X}\right)\right), \\
    &\sigma_{\text{FITC}*}^2 = \kappa_{**} - \mathbf{q}_{*y} \mathbf{K}_{\text{FITC}}^{-1} \mathbf{q}_{y*},
\end{align}
where $\mu_{\text{FITC}*}$ and $\sigma_{\text{FITC}*}^2$ are posterior predictive mean and variance, respectively, and $\kappa_{**}$ denotes the self-covariance at test point ($\kappa\left(\mathbf{x}_*, \mathbf{x}_*\right)$).

\section{Design Coupling Estimation}
\label{sec:DC_est}

\begin{figure*}
\begin{align}\label{eqn:dcm_2}
\mathbf{J}_{\mathbf{x}} = \left[
    \begin{array}{ccc}
        \left|\left|\left(
            \left.
                \dfrac{\partial \hat{x}_1^*}{\partial \hat{x}_1}
            \right|_{\text{pert}_1}, \cdots,
            \left.
                \dfrac{\partial \hat{x}_1^*}{\partial \hat{x}_1}
            \right|_{\text{pert}_{N_s}}
        \right)^{\textsf{T}}\right|\right|
        & \cdots
        & \left|\left|\left(
            \left.
                \dfrac{\partial \hat{x}_1^*}{\partial \hat{x}_{N_x}}
            \right|_{\text{pert}_1}, \cdots,
            \left.
                \dfrac{\partial \hat{x}_1^*}{\partial \hat{x}_{N_x}}
            \right|_{\text{pert}_{N_s}}
        \right)^{\textsf{T}}\right|\right| \\
        \vdots & \ddots & \vdots \\
        \left|\left|\left(
            \left.
                \dfrac{\partial \hat{x}_{N_x}^*}{\partial \hat{x}_1}
            \right|_{\text{pert}_1}, \cdots,
            \left.
                \dfrac{\partial \hat{x}_{N_x}^*}{\partial \hat{x}_1}
            \right|_{\text{pert}_{N_s}}
        \right)^{\textsf{T}}\right|\right|
        & \cdots
        & \left|\left|\left(
            \left.
                \dfrac{\partial \hat{x}_{N_x}^*}{\partial \hat{x}_{N_x}}
            \right|_{\text{pert}_1}, \cdots,
            \left.
                \dfrac{\partial \hat{x}_{N_x}^*}{\partial \hat{x}_{N_x}}
            \right|_{\text{pert}_{N_s}}
        \right)^{\textsf{T}}\right|\right|
    \end{array}
\right],
\end{align}
\end{figure*}
\begin{figure*}
\begin{align}\label{eqn:osm_2}
\mathbf{J}_{\bm{\Psi}} = \left[
    \begin{array}{ccc}
        \left|\left|\left(
            \left.
                \dfrac{\partial \hat{\Psi}(\hat{x}_1^{*})}{\partial \hat{x}_1}
            \right|_{\text{pert}_1}, \cdots,
            \left.
                \dfrac{\partial \hat{\Psi}(\hat{x}_1^{*})}{\partial \hat{x}_1}
            \right|_{\text{pert}_{N_s}}
        \right)^{\textsf{T}}\right|\right|
        & \cdots
        & \left|\left|\left(
            \left.
                \dfrac{\partial \hat{\Psi}(\hat{x}_1^{*})}{\partial \hat{x}_{N_x}}
            \right|_{\text{pert}_1}, \cdots,
            \left.
                \dfrac{\partial \hat{\Psi}(\hat{x}_1^{*})}{\partial \hat{x}_{N_x}}
            \right|_{\text{pert}_{N_s}}
        \right)^{\textsf{T}}\right|\right| \\
        \vdots & \ddots & \vdots \\
        \left|\left|\left(
            \left.
                \dfrac{\partial \hat{\Psi}(\hat{x}_{N_x}^{*})}{\partial \hat{x}_1}
            \right|_{\text{pert}_1}, \cdots,
            \left.
                \dfrac{\partial \hat{\Psi}(\hat{x}_{N_x}^{*})}{\partial \hat{x}_1}
            \right|_{\text{pert}_{N_s}}
        \right)^{\textsf{T}}\right|\right|
        & \cdots
        & \left|\left|\left(
            \left.
                \dfrac{\partial \hat{\Psi}(\hat{x}_{N_x}^{*})}{\partial \hat{x}_{N_x}}
            \right|_{\text{pert}_1}, \cdots,
            \left.
                \dfrac{\partial \hat{\Psi}(\hat{x}_{N_x}^{*})}{\partial \hat{x}_{N_x}}
            \right|_{\text{pert}_{N_s}}
        \right)^{\textsf{T}}\right|\right|
    \end{array}
\right],
\end{align}
\end{figure*}

The use of surrogate model with finite-dimensional input and output quantities eliminates the need to repeatedly evaluate the full system dynamics during the solution of the system design optimization problems. Given that the constructed models have the finite-dimensional structure (as opposed to functions in continuous time-domain spaces), the same DCA methods employed for MDO problems in the authors' earlier work, such as Chinthoju, et al.~\cite{Chinthoju2024eVTOL} and Fern{\'a}ndez~Bravo and Allison~\cite{FernandezBravo2024SatelliteMDO}, are directly applicable. Design coupling estimated in this work follows Approach 2 presented in Fern{\'a}ndez~Bravo and Allison~\cite{FernandezBravo2024SatelliteMDO}, and this is further explained below. Two Jacobian matrices are used to characterize design coupling strength: one for the optimal design variable sensitivities (herein referred to as \emph{design coupling matrix}) and another for the sensitivities of the optimal objective function values (referred to as \emph{objective sensitivity matrix}). The Jacobian matrices are obtained by the following procedures:
\begin{enumerate}
    \item A sweep of $N_s$ optimizations is conducted. In each optimization  problem, one variable is selected as the optimization variable, a second variable is perturbed across its sweep range (from lower to upper bounds), and all remaining variables are fixed at their nominal values. This process is repeated with all combinations of optimization and perturbed variables.
    \item After all the optimization sweeps are performed, the Jacobian matrices are calculated. For each point during the sweep, two Jacobian matrices are obtained, resulting in multiple sets of design coupling and objective sensitivity matrices.
    \item To obtain a single set of Jacobian matrices from the sweeps, we calculate the element-wise norm of each entry in the Jacobian matrices across all points in the sweep.
\end{enumerate}

The design coupling matrix ($\mathbf{J}_{\mathbf{x}}$) is defined in Eq.~\eqref{eqn:dcm_2}, and the objective sensitivity matrix ($\mathbf{J}_{\bm{\Psi}}$) is given in Eq.~\eqref{eqn:osm_2}. In both formulations, sensitivities are expressed in Jacobian form.

The design coupling matrix characterizes interdependence between distinct design variables and describes how the optimal value of one variable varies in response to perturbations in another as the design space is explored. Each Jacobian entry, $\partial \hat{x}_A^* / \partial \hat{x}_B^{}$, represents the sensitivity of the optimized value of design variable $A$ with respect to a perturbation in variable $B$. This sensitivity is computed by re-optimizing variable $A$ while perturbing variable $B$, with all design variables fixed at their nominal values. The perturbation of variable $B$ is evaluated at $N_s$ discrete points, uniformly distributed between its lower and upper bounds, denoted as $\text{pert}_i$ where $i \in [1, N_s]$.

The objective sensitivity matrix quantifies how the optimal objective function value varies under the same DCA, i.e., when one design variable is perturbed and another is re-optimized accordingly. Its entries correspond to the sensitivities of the objective function values obtained during the construction of the design coupling matrix. Each corresponding Jacobian entry is given by $\partial \hat{\Psi}(\hat{x}_A^{*})/\partial \hat{x}_B^{}$, representing the sensitivity of the optimized objective function value with respect to perturbations in variable $B$, while variable $A$ is re-optimized and all remaining variables are fixed at their nominal values.

All input and output variables are normalized in every approach, as indicated by the $\hat{\;\;}$ (hat) notation. This normalization is particularly important when dealing with design variables and response quantities that span different scales. By placing all variables on a consistent scale range, normalization ensures meaningful sensitivity evaluation and enables fair comparison across different design variables and performance metrics.

This approach provides sensitivity information that is valid over a representative slice of the design space by systematically exploring each variable across its full range, from lower to upper bounds. By doing so, it captures the direct influence of perturbing one design variable on the optimal value of another.

\section{Problem Definition}\label{sec:fowt_prob}

This article presents case studies based on the IEA 15~MW wind turbine and the VolturnUS-S floating platform, as described in Sect.~\ref{sec:baselinewtdesign}, and implemented using the WEIS toolset described in Sect.~\ref{sec:weistoolset}.

The objective function across all case studies is to minimize the platform mass, $m_{\text{ptfm}}$, which is the total mass of all constituent members modeled within the RAFT input model, all within the WEIS framework. Two nonlinear constraints are imposed within the problem, enforcing maximum platform pitching displacement, $\max(\theta_{\text{ptfm}}) \le \theta_{\text{ptfm,limit}} = 6.9^{\circ}$, and maximum nacelle acceleration, $\max(a_{\text{nac,f-a}}) \le a_{\text{nac,f-a,limit}} = 0.7$ m/s$^2$, both over all design load cases (DLCs) evaluated within the design optimization loop.

\begin{figure}[t]
    \centering
    \includegraphics[width=0.8\linewidth]{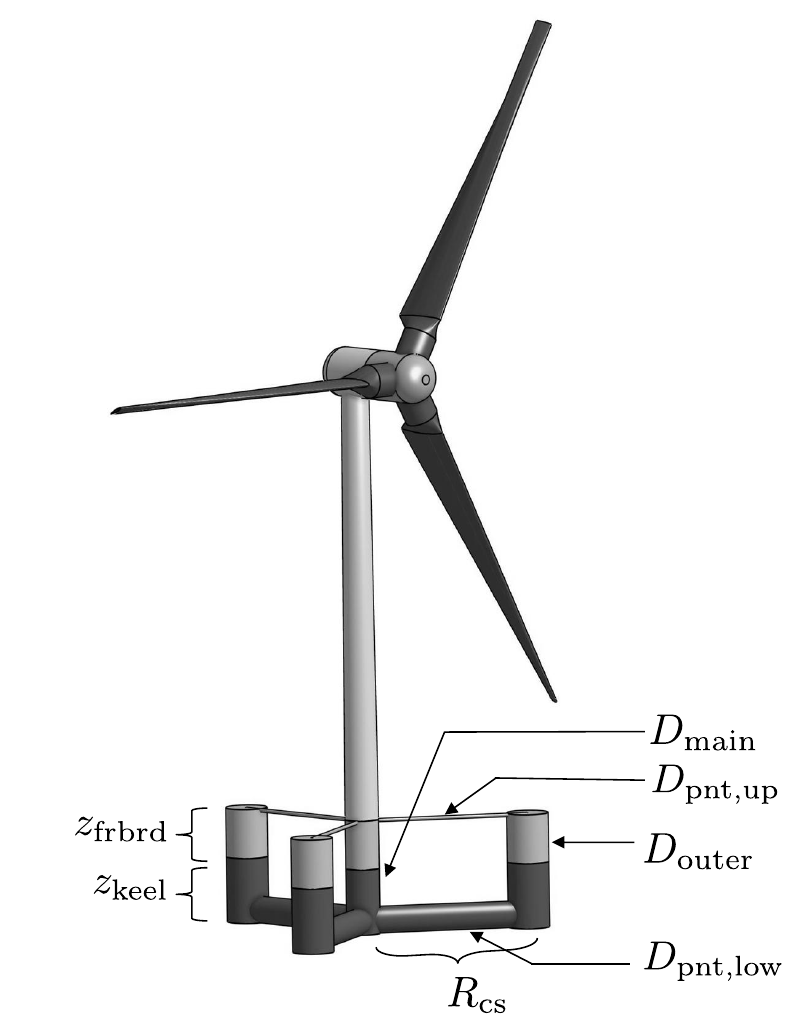}
    \caption{Floating offshore wind turbine model with annotated plant design variables used in this study. Column sections in light shading depict the freeboard (portions situated above the still-water level (SWL)), whereas dark-shaded sections indicate the portions that remain submerged below the SWL.}
    \label{fig:turbine} 
\end{figure}

Seven plant design variables, selected from the floating platform geometric parameters, are considered in this study: the diameter of lower pontoon, $D_{\text{pnt,low}}$; the keel depth in the $z$-coordinate, $z_{\text{keel}}$; the diameter of the upper pontoon, $D_{\text{pnt,up}}$; the diameter of the main column, $D_{\text{main}}$; the diameter of the outer column, $D_{\text{outer}}$, the freeboard height above the water surface in the $z$-coordinate, $z_{\text{frbrd}}$; and the spacing from the center column to the outer columns, measured from centerline to centerline of each cylindrical column, $R_{\text{cs}}$. One control variable is also considered in the studies: the percent peak shaving, $\textsf{ps}_{\text{\%}}$. The peak shaving defines a minimum blade pitch schedule near rated operation \cite{ABBAS2024ccd}. The candidate variables and their nominal values and bounds are shown in Table~\ref{tab:FOWT_dvs}. Figure~\ref{fig:turbine} illustrates the FOWT geometry with the plant design variables.

\begin{table}[t]
    \caption{Floating offshore wind turbine optimization problem candidate design variables.}\label{tab:FOWT_dvs}%
    \centering{%
    \begin{tabular}{@{}lrrrc@{}}
        \toprule
        & \multicolumn{2}{c}{Bounds} & Nominal & \\
        \cmidrule(lr){2-3}
        Variable & Lower$\;\;$ & Upper$\;\;$ & value$\;\;\;$ & Unit  \\
        \midrule
        $D_{\text{main}}$ & 6.0000 & 14.0000 & 10.0000 & $m$\\
        $D_{\text{pnt,up}}$ & 0.7100 & 1.1100 & 0.9100 & $m$ \\
        $D_{\text{pnt,low}}$ & 6.6148 & 13.6148 & 9.6148 & $m$\\
        $D_{\text{outer}}$ & 10.5000 & 14.5000 & 12.5000 & $m$ \\
        $R_{\text{cs}}$ & 41.5700 & 61.5700 & 51.7500 & $m$  \\
        $z_{\text{keel}}$ & $-$24.0000 & $-$16.0000 & $-$20.0000 & $m$  \\
        $z_{\text{frbrd}}$ & 7.0000 & 21.0000 & 15.0000 & $m$  \\
        $\textsf{ps}_{\text{\%}}$ & 0.7500 & 1.0000 & 0.8500 & - \\
        \bottomrule
    \end{tabular}
    }%
\end{table}

Due to disciplinary couplings, as well as the number and complexity of the associated discipline models, the multidisciplinary design analysis (MDA) of the FOWT system is computationally demanding. Furthermore, because the DCA requires solving a large number of optimization problems across comprehensive combinations of design variable perturbations, it becomes computationally intractable, particularly as the number of design variables increases, even with the lower-fidelity models, such as RAFT. Thus, we conducted DCA with sensitivities obtained through the constructed surrogate model on the responses of the RAFT model.

The surrogate was trained using 750 samples generated through Latin hypercube sampling and implemented using SGP modeling, as described in Sect.~\ref{sec:sgp}. The SGP surrogate was developed through customized code built upon the open-source surrogate modeling toolbox (SMT) python module~\cite{savesSMT20Surrogate2024}. The constructed surrogate model takes plant and control design variables, including $\textsf{ps}_{\%}$, $D_{\text{pnt,low}}$, $z_{\text{keel}}$, $D_{\text{pnt,up}}$, $D_{\text{main}}$, $D_{\text{outer}}$, $z_{\text{frbrd}}$, and $R_{\text{cs}}$, plus six additional control parameters (although not used in this study) as input parameters. There are 46 parameters provided by the trained surrogate model as outputs, which characterize the response and performance of the FOWT system of interest. The output parameters include the platform mass, $m_{\text{ptfm}}$, the maximum platform pitching displacement, $\max(\theta_{\text{ptfm}})$, and the maximum nacelle fore-aft acceleration, $\max(a_{\text{nac,f-a}})$.

\section{Results}
\label{sec:results}

\begin{figure}[t!]
\centering\includegraphics[width=0.95\linewidth]{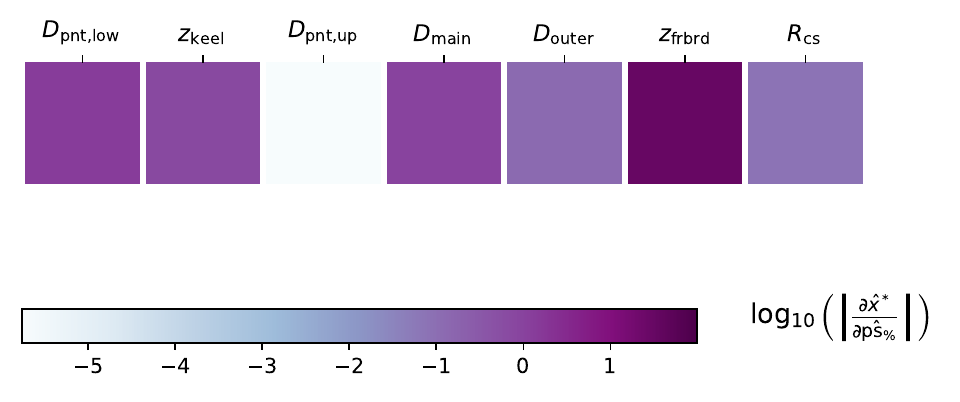}
\caption{Heat map representing the control-plant coupling.}
\label{fig:dxdu_fowt} 
\end{figure}

\begin{figure}[t!]
\centering\includegraphics[width=0.95\linewidth]{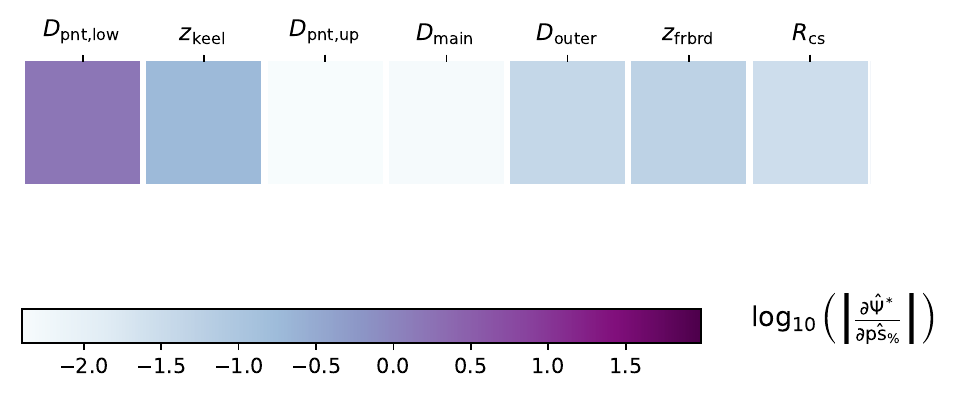}
\caption{Heat map representing the sensitivity of the optimal objective function value with respect to control changes.}
\label{fig:dfdu_fowt} 
\end{figure}

\begin{figure}[t!]
\centering\includegraphics[width=0.95\linewidth]{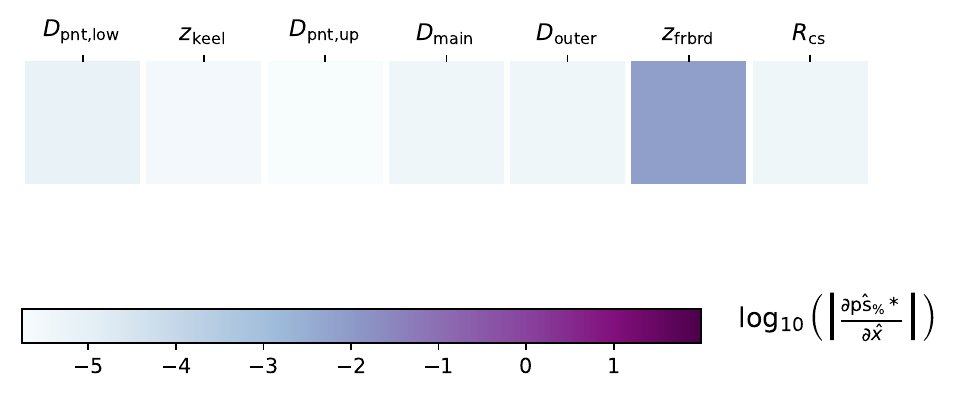}
\caption{Heat map representing the plant-control coupling.}
\label{fig:dudxp_fowt} 
\end{figure}

This section presents the results of the DCA and demonstrates how the obtained information can be leveraged to support FOWT optimization. Section~\ref{sec:design-coupling-estimation} introduces the estimation of design coupling, including detailed evaluations of control-plant, plant-control, and plant-plant couplings. Based on these coupling characteristics, two case studies are presented. Section~\ref{sec:result-case-sequence} (Case Study 1) uses the design coupling information to decompose the overall problem into a sequence of smaller optimization subproblems. Section~\ref{sec:result-case-subsetoptim} (Case Study 2) employs the coupling results to identify the most impactful set of design variables for the optimization problem.

\subsection{Design Coupling Estimation} \label{sec:design-coupling-estimation}

Estimations of the bidirectional coupling between control and plant design variables, as well as the coupling among plant design variables, are presented in this section. The detailed methodology used to compute the design coupling and objective sensitivity matrices for these coupling estimations is provided in Sec.~\ref{sec:DC_est}.

\subsubsection{Control-Plant Coupling}

Figure~\ref{fig:dxdu_fowt} presents the influence of changes in control design on the optimal plant design. In this figure, stronger coupling is indicated by dark shade and weaker coupling by light shade. The results indicate that all plant design variables, with the exception of $D_{\text{pnt,up}}$, have significant degrees of coupling with the control design decision. The floating platform's freeboard height above the water surface in the $z$-coordinate ($z_{\text{frbrd}}$) exhibits the highest sensitivity to changes in control design decision.

Figure~\ref{fig:dfdu_fowt} illustrates the effect of the same control-plant coupling analysis on the objective function value. In this figure, stronger influence is indicated by dark shade and weaker influence by light shade. In this study, the control design variable ($\textsf{ps}_{\%}$) is perturbed, and the plant design variables are subsequently re-optimized to evaluate the resulting change in the objective function. The influence of $\textsf{ps}_{\%}$ on the objective function value is relatively small, and it is even negligible when the optimization is performed directly using the WEIS model. This discrepancy is likely due to local approximation errors in the surrogate model.

\subsubsection{Plant-Control Coupling}

Figure~\ref{fig:dudxp_fowt} presents the influence of changes in plant design on the optimal control design. In this figure, stronger coupling is indicated by dark shade and weaker coupling by light shade. The results indicate that, with the exception of $z_{\text{frbrd}}$, the plant design variables have a negligible effect on the optimal control design solution. Even in the case of the floating platform's freeboard height above the water surface in the $z$-coordinate, the observed coupling to the control variable remains weak.

Considered together with the control-plant coupling results, these findings suggest that the coupling between plant and control is predominantly unidirectional, with the optimal control design being largely independent of the plant design decisions. Consequently, this problem is well suited for a sequential optimization strategy.

When the problem is solved using a simultaneous optimization approach, the resulting objective function value is $7.741 \times 10^6 \; \text{kg}$, which we adopt as the best attainable design for the studies presented in this paper. A very similar objective function value of $8.007 \times 10^6 \; \text{kg}$ is obtained using the sequential optimization approach. The close agreement between these results confirms that the sequential solution strategy suggested by the DCA is appropriate approach for this problem.

\subsubsection{Plant-Plant Coupling}

\begin{figure}[t]
\centering\includegraphics[width=0.95\linewidth]{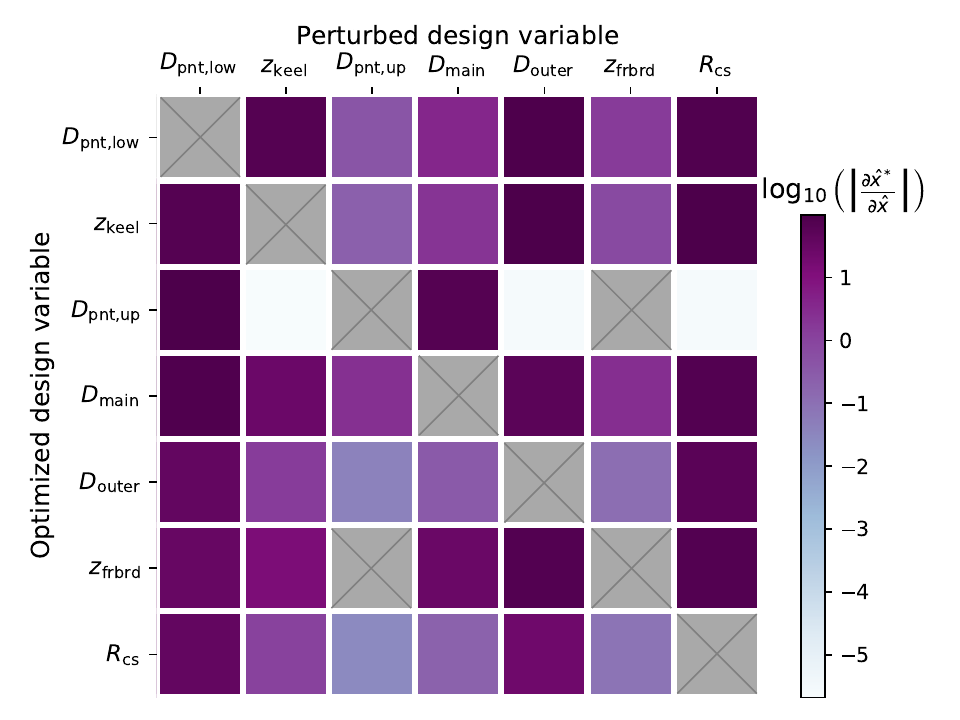}
\caption{Heat map representing the design coupling matrix.}
\label{fig:dxdx_mat} 
\end{figure}

\begin{figure}[t]
\centering\includegraphics[width=0.95\linewidth]{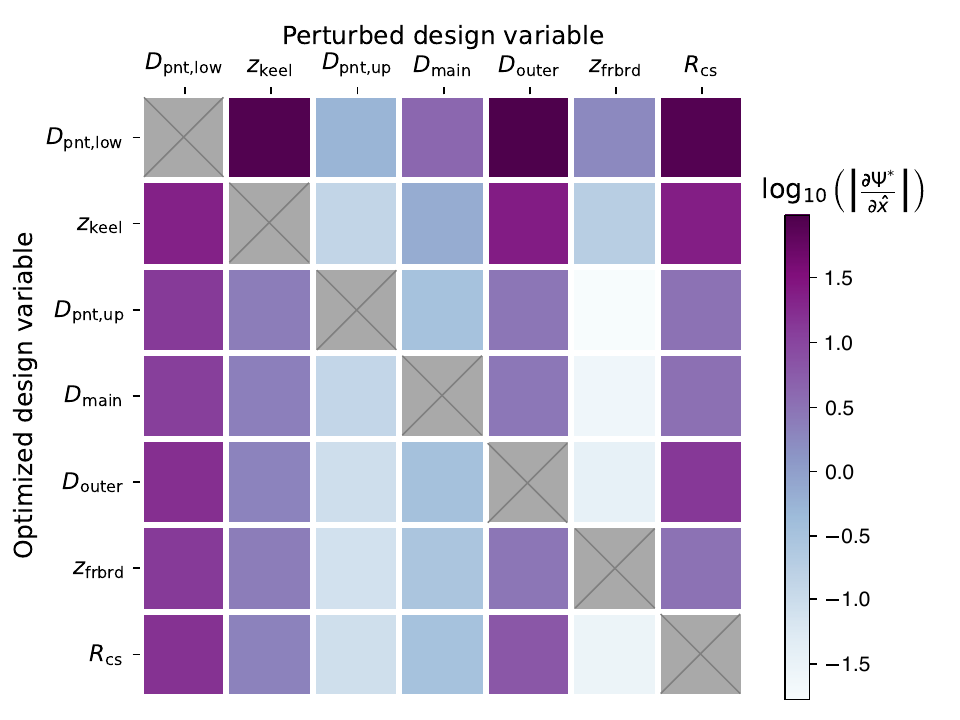}
\caption{Heat map representing the objective sensitivity matrix.}
\label{fig:dfdx_mat} 
\end{figure}

In this section, the couplings among plant design variables are analyzed. The design coupling matrix and the objective sensitivity matrix are presented in Figs.~\ref{fig:dxdx_mat} and~\ref{fig:dfdx_mat}, respectively. In these figures, stronger coupling is indicated by dark shade, weaker coupling by light shade, and negligible coupling (off-diagonals) or not defined (diagonals) by shade with cross mark.

The design coupling matrix in Fig.~\ref{fig:dxdx_mat} reveals the directional coupling strengths for all pairwise combinations of plant design variables. This interaction is bidirectional. Although highly symmetric trends can be observed, the resulting design coupling matrix is not perfectly symmetric. This indicates that the coupling strength observed by perturbing design variable $A$ and optimizing design variable $B$ does not exactly match the strength observed by perturbing design variable $B$ and optimizing design variable $A$. The strongest coupling is observed between $D_{\text{pnt,low}}$ and $D_{\text{main}}$.

Based on Fig.~\ref{fig:dfdx_mat}, the plant design variables ranked from most to least influential on the objective function value are: $D_{\text{pnt,low}}$, $z_{\text{keel}}$, $D_{\text{outer}}$, $R_{\text{cs}}$, $z_{\text{frbrd}}$, $D_{\text{pnt,up}}$, and $D_{\text{main}}$. Among these variables, $D_{\text{pnt,low}}$ has a substantially greater impact on the objective function value than any other variable. However, in contrast, the influences of $z_{\text{frbrd}}$, $D_{\text{pnt,up}}$, and $D_{\text{main}}$ on the objective function value are relatively small.

\subsection{Case Study 1: Sequential Decomposition Based on Design Coupling} \label{sec:result-case-sequence}

The goal of this case study is to reformulate the full simultaneous optimization problem into a sequence of compact optimization subproblems that produces a design solution close to the optimum from the simultaneous design problem while reducing computational efforts, and demonstrate the usefulness and effectiveness of the design coupling information. The decomposition method leverages the design coupling matrix to group design variables and construct design sequence of them. Beyond improving computational efficiency, this approach also serves to validate the qualitative insights derived from the DCA and to provide deeper insights into the underlying design mechanisms.

The process begins with an interpretation of the design coupling Jacobian matrix, visualized in Fig.~\ref{fig:dxdx_mat}. In this plot, darker shaded cells indicate strong coupling between two design variables, whereas lighter cells represent weaker coupling. The matrix is not perfectly symmetric about the diagonal, due to the presence of directional couplings. For certain variable pairs, the influence of one variable on the optimal solution of the other is significantly stronger than the reverse relationship.

Using this information, a decision influence chain is constructed and translated into a sequence of optimization subproblems. Design variables that strongly influence the optimal decision of other variables, while being weakly influenced in the reverse relationship, are prioritized and optimized in earlier stages of the sequence. Conversely, variables that have optimal decisions strongly affected by others but exert limited influence to the decision of others are deferred to later stages. Variables exhibiting strong bidirectional coupling are grouped and optimized simultaneously within the same subproblem to preserve their interdependence.

\begin{figure}[t]
    \centering
    \includegraphics[width=0.4\linewidth]{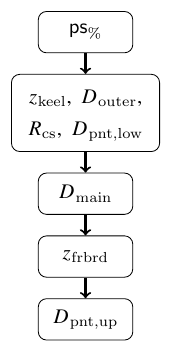}
    \caption{Optimization sequence generated based on information from design coupling {J}acobian matrix.}
    \label{fig:sequence}
\end{figure}

This decomposition strategy, guided by the DCA, transforms the original large-scale problem into a set of smaller optimization subproblems. It enhances computational tractability while retaining the dominant coupling relationships that govern the optimal design decisions. The resulting sequence constructed through this process is given in Fig.~\ref{fig:sequence}.

The constructed sequence starts with the optimization of $\textsf{ps}_{\text{\%}}$, since its optimal decision is weakly influenced by change of decisions in other variables. In this first stage, the original optimization formulation is retained, but all design variables except $\textsf{ps}_{\text{\%}}$ are held fixed at their nominal values. The objective function is to minimize the platform mass, $m_{\text{ptfm}}$, subject to constraints on the maximum platform pitching displacement,$\max(\theta_{\text{ptfm}}) \le \theta_{\text{ptfm,limit}}$, and maximum nacelle acceleration, $\max(a_{\text{nac,f-a}}(a)) \le a_{\text{nac,f-a,limit}}$. The resulting problem formulation for the first stage of the sequence is given as:
\begingroup
\renewcommand{\arraystretch}{1.5}
\begin{align}
    \allowdisplaybreaks
    \begin{array}{rl}
        \text{minimize:}
            & m_{\text{ptfm}} \\
        \text{with respect to:}
            & \mathbf{x}_{\text{seq1}} = [\textsf{ps}_{\%}]^{\textsf{T}} \\
        \text{subjet to:}
            & \max(\theta_{\text{ptfm}}) \le \theta_{\text{ptfm,limit}} \\
            & \max(a_{\text{nac,f-a}}) \le a_{\text{nac,f-a,limit}}
    \end{array}
\end{align}
\endgroup
where
\begingroup
\renewcommand{\arraystretch}{1.5}
\begin{align*}
    \allowdisplaybreaks
    \begin{array}{l}
        D_{\text{pnt,low}} \leftarrow D_{\text{pnt,low}}^{0} \\
        z_{\text{keel}} \leftarrow z_{\text{keel}}^{0} \\
        D_{\text{pnt,up}} \leftarrow D_{\text{pnt,up}}^{0} \\
        D_{\text{main}} \leftarrow D_{\text{main}}^{0} \\
        D_{\text{outer}} \leftarrow D_{\text{outer}}^{0} \\
        z_{\text{frbrd}} \leftarrow z_{\text{frbrd}}^{0} \\
        R_{\text{cs}} \leftarrow R_{\text{cs}}^{0} \\
        \mathbf{x} = \left[
            D_{\text{pnt,low}},
            z_{\text{keel}},
            D_{\text{pnt,up}},
            D_{\text{main}},
            D_{\text{outer}},
            z_{\text{frbrd}},
            R_{\text{cs}},
            \textsf{ps}_{\%}
        \right]^{\textsf{T}} \\
        \left[
            m_{\text{ptfm}},
            \max(\theta_{\text{ptfm}}),
            \max(a_{\text{nac,f-a}})
        \right]^{\textsf{T}} \leftarrow \textsf{WEIS}(\mathbf{x})
    \end{array}
\end{align*}
\endgroup

In the next stage, the four variables $z_{\text{keel}}$, $D_{\text{outer}}$, $R_{\text{cs}}$, and $D_{\text{pnt,low}}$ are optimized simultaneously, as they exhibit strong bidirectional coupling among themselves. The design decision for $\textsf{ps}_{\%}$ strongly influences the optimal decisions of these variables, and the design decision of these four variables significantly affect the optimal decisions of most of the remaining design variables. Therefore, this group is prioritized immediately after the control design variable subproblem in the optimization sequence. The variable $\textsf{ps}_{\%}$ is updated to the optimal solution obtained in the previous stage and held fixed during this stage of the sequence. The resulting problem formulation for the second stage of the sequence is given as:
\begingroup
\renewcommand{\arraystretch}{1.5}
\begin{align}
    \allowdisplaybreaks
    \begin{array}{rl}
        \text{minimize:}
            & m_{\text{ptfm}} \\
        \text{with respect to:}
            & \mathbf{x}_{\text{seq2}} = [
                z_{\text{keel}},
                D_{\text{outer}},
                R_{\text{cs}},
                D_{\text{pnt,low}}
            ]^{\textsf{T}} \\
        \text{subjet to:}
            & \max(\theta_{\text{ptfm}}) \le \theta_{\text{ptfm,limit}} \\
            & \max(a_{\text{nac,f-a}}) \le a_{\text{nac,f-a,limit}}
    \end{array}
\end{align}
\endgroup
where
\begingroup
\renewcommand{\arraystretch}{1.5}
\begin{align*}
    \allowdisplaybreaks
    \begin{array}{l}
        \textsf{ps}_{\%} \leftarrow \textsf{ps}_{\%}^{\ast} \\
        D_{\text{pnt,up}} \leftarrow D_{\text{pnt,up}}^{0} \\
        D_{\text{main}} \leftarrow D_{\text{main}}^{0} \\
        z_{\text{frbrd}} \leftarrow z_{\text{frbrd}}^{0} \\
        \mathbf{x} = \left[
            D_{\text{pnt,low}},
            z_{\text{keel}},
            D_{\text{pnt,up}},
            D_{\text{main}},
            D_{\text{outer}},
            z_{\text{frbrd}},
            R_{\text{cs}},
            \textsf{ps}_{\%}
        \right]^{\textsf{T}} \\
        \left[
            m_{\text{ptfm}},
            \max(\theta_{\text{ptfm}}),
            \max(a_{\text{nac,f-a}})
        \right]^{\textsf{T}} \leftarrow \textsf{WEIS}(\mathbf{x})
    \end{array}
\end{align*}
\endgroup

Similarly, the remaining three design variables are sequentially optimized in the order of $D_{\text{main}}$, $z_{\text{frbrd}}$, and $D_{\text{pnt,up}}$. This order is based on the directional coupling information obtained through the design coupling matrix. At each stage, the optimal values obtained from the preceding steps are updated and held fixed while optimizing the current design variable.

\begin{figure}[t]
\centering
\includegraphics[width=0.985\linewidth]{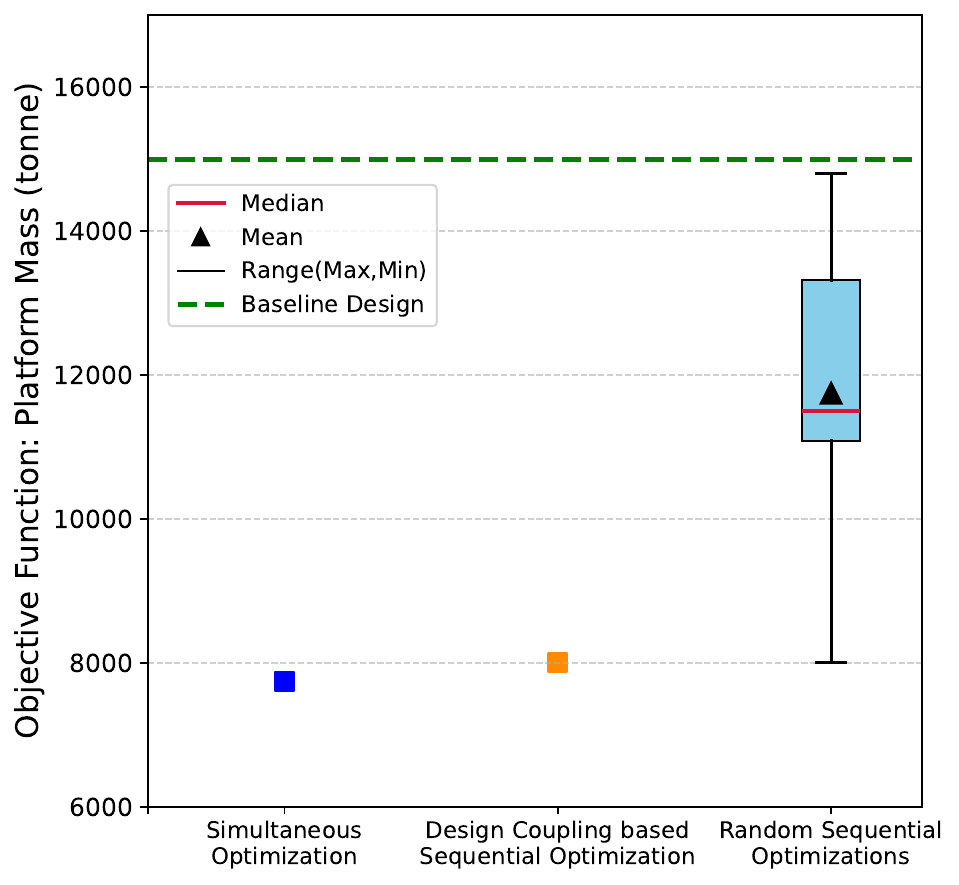}
\caption{Comparison of objective function values by optimization strategies.}
\label{fig:strategy_compar}
\end{figure}

The results obtained from the proposed sequential decomposition method based on design coupling are compared to the optimal solution from the full simultaneous optimization, as well as 64 unique randomly-generated sequential optimization cases. Each random case follows the same sequential structure shown in Fig.~\ref{fig:sequence}: one variable is optimized in the first stage, four variables in the second stage, and a single variable in each of the third, fourth, and fifth stages, but with a randomly assigned ordering of variables within this framework.

\begin{table*}[t]
\caption{Results with random sequences and proposed DCA-based sequence.}
\label{tab:sequence_results}
\centering
\begin{tabular}{@{}c c c c c c@{}}
\toprule
1st sequence & 2nd sequence & 3rd sequence & 4th sequence & 5th sequence & $m_{\text{ptfm}}^*$ (kg)\\
\midrule
$\textsf{ps}_{\text{\%}}$ & $D_{\text{pnt,low}}$, $D_{\text{outer}}$, $R_{\text{cs}}$, $z_{\text{keel}}$ & $D_{\text{main}}$ & $z_{\text{frbrd}}$ & $D_{\text{pnt,up}}$ & (Proposed Sequence) $8.006770 {\times} 10^{6}$\\
$D_{\text{pnt,up}}$ & $D_{\text{pnt,low}}$, $D_{\text{outer}}$, $R_{\text{cs}}$, $z_{\text{keel}}$ & $\textsf{ps}_{\text{\%}}$ & $z_{\text{frbrd}}$ & $D_{\text{main}}$ & $8.006770 {\times} 10^{6}$ \\
$z_{\text{keel}}$ & $D_{\text{pnt,low}}$, $D_{\text{outer}}$, $R_{\text{cs}}$, $z_{\text{frbrd}}$ & $\textsf{ps}_{\text{\%}}$ & $D_{\text{pnt,up}}$ & $D_{\text{main}}$ & $8.583204 {\times} 10^{6}$\\
$z_{\text{keel}}$ & $D_{\text{pnt,low}}$, $D_{\text{outer}}$, $R_{\text{cs}}$, $\textsf{ps}_{\text{\%}}$ & $z_{\text{frbrd}}$ & $D_{\text{pnt,up}}$ & $D_{\text{main}}$ & $8.610103 {\times} 10^{6}$ \\
$z_{\text{keel}}$ & $D_{\text{pnt,low}}$, $D_{\text{outer}}$, $R_{\text{cs}}$, $\textsf{ps}_{\text{\%}}$ & $z_{\text{frbrd}}$ & $D_{\text{pnt,up}}$ & $D_{\text{main}}$ & $8.610103{\times} 10^{6}$ \\
$\textsf{ps}_{\text{\%}}$ & $D_{\text{pnt,low}}$, $D_{\text{outer}}$, $R_{\text{cs}}$, $D_{\text{main}}$ & $z_{\text{frbrd}}$ & $z_{\text{keel}}$ & $D_{\text{pnt,up}}$ & $8.794438 {\times} 10^{6}$ \\
$D_{\text{pnt,up}}$ & $D_{\text{pnt,low}}$, $D_{\text{outer}}$, $R_{\text{cs}}$, $D_{\text{main}}$ & $z_{\text{frbrd}}$ & $\textsf{ps}_{\text{\%}}$ & $z_{\text{keel}}$ & $8.794438 {\times} 10^{6}$ \\
\bottomrule
\end{tabular}
\end{table*}

The overall comparison of the results is presented in Fig.~\ref{fig:strategy_compar}. The vertical axis represents the optimal objective function value (i.e., platform mass, scaled in tonne) achieved by each method, while the horizontal axis identifies the corresponding optimization strategies. The dashed horizontal line denotes the objective function value of the baseline design ($1.509260 \times 10^{7}$ kg). The results from the 64 randomly generated sequential optimization cases are illustrated using a box plot. In this representation, the triangle symbol indicates the mean value, the solid horizontal line within the box represents the median, and the whiskers denote the minimum and maximum values.

The design solution obtained by the sequential decomposition method based on design coupling analysis converged to an objective function value of $8.006770 \times 10^{6}$ kg, which is only 3.4\% higher than the optimal solution obtained from the full simultaneous optimization, $7.740970 \times 10^{6}$ kg. In contrast, the objective function values produced by the randomly-generated sequential optimization cases range from as high as $1.479707 \times 10^{7}$ to as low as $8.006770 \times 10^{6}$, which is equivalent to the result of the sequential decomposition approach. The mean objective function value across the random sequences is $1.174689 \times 10^{7}$ kg.

A closer look at the results from the random cases indicates that the second stage of the sequence, where four variables are optimized simultaneously, has the most significant influence on the final optimal solution. In particular, random cases that grouped $z_{\text{keel}}$, $D_{\text{outer}}$, $R_{\text{cs}}$, and $D_{\text{pnt,low}}$ for simultaneous optimization resulted significantly lower objective function values compared to cases with other variable combinations. Notably, the lowest objective function value among the random cases is numerically identical to the result from the proposed sequential decomposition method based on design coupling, strongly supporting the underlying hypothesis regarding the importance of preserving dominant bidirectional couplings. These results are summarized in Table~\ref{tab:sequence_results}.

\subsection{Case Study 2: Identification of Influential Variable Sets Based on Design Coupling} \label{sec:result-case-subsetoptim}

Case Study 1 presented in Sec.~\ref{sec:result-case-sequence} demonstrated that certain combinations of design variables, when optimized simultaneously, have a significant impact on the final objective function value. Building on this observation, the goal of this case study is to systematically identify the most influential sets of design variables for the optimization problem. This identification leverages both the design coupling matrix, given in Fig.~\ref{fig:dxdx_mat}, which captures the interdependencies among design variables, and the objective sensitivity matrix, given in Fig.~\ref{fig:dfdx_mat}, which quantifies the direct impact of pairs of variable changes on the objective function value.

The study begins with the interpretation of the information contained in the objective sensitivity matrix. When selecting a single plant design variable, the decision is based solely on its direct influence on the objective function value given that there is no coupling that the optimizer can exploit to further improve the objective function value. The diameter of lower pontoon, $D_{\text{pnt,low}}$, is the most prominent candidate due to its largest objective sensitivity magnitude. Consistent with this observation, optimizing the FOWT model using WEIS with only this single design variable achieves the lowest objective function value among all single plant variable cases, resulting in a platform mass of $1.166550 \times 10^{7}$ kg. The corresponding results are summarized in Table~\ref{tab:weis_1dv}.

\begin{table}[t]
    \centering
    \caption{Optimization results using {WEIS} with one plant design variable.}\label{tab:weis_1dv}%
    \begin{tabular}{@{}cc@{}}
        \toprule
         Design variable&   $m_{\text{ptfm}}^*$ (kg)\\
         \midrule
         $D_{\text{pnt,low}}$ &$1.166550{\times} 10^{7}$\\
         $z_{\text{keel}}$ & $1.386380{\times} 10^{7}$\\
         $D_{\text{outer}}$& $1.492430{\times} 10^{7}$ \\
         $R_{\text{cs}}$ & $1.493140{\times} 10^{7}$ \\
         $z_{\text{frbrd}}$& $1.509260{\times} 10^{7}$\\
         $D_{\text{pnt,up}}$& $1.509260{\times} 10^{7}$\\
         $D_{\text{main}}$ &$1.509260{\times} 10^{7}$ \\
        \bottomrule
    \end{tabular}
\end{table}

When selecting two design variables for optimization, multiple selection strategies may be considered. The first approach selects the two variables with the highest direct impact on the objective function values, namely $D_{\text{pnt,low}}$ and $z_{\text{keel}}$. The second approach begins with selecting the most impactful design variable, $D_{\text{pnt,low}}$, and pairs it with the design variable most strongly coupled to it according to the design coupling matrix, which in this study corresponds to $D_{\text{main}}$.

\begin{table*}[t]
\caption{Optimization results using {WEIS-RAFT} for a sampling of design variable combinations and objective function value for the same designs with {WEIS-OpenFast}.}\label{tab:weis}%
\centering{%
\begin{tabular}{@{}lrr|lrr@{}}
\toprule
Design Variables & \begin{tabular}{@{}c@{}} $m_{\text{ptfm}}^*$ (kg) \\ (RAFT)\end{tabular}  & \begin{tabular}{@{}c@{}} $m_{\text{ptfm}}$ (kg)\\ (OpenFAST)\end{tabular} & Design Variables & \begin{tabular}{@{}c@{}} $m_{\text{ptfm}}^*$ (kg)\\ (RAFT)\end{tabular} & \begin{tabular}{@{}c@{}} $m_{\text{ptfm}}$ (kg)\\ (OpenFAST)\end{tabular}\\ \midrule
$D_{\text{pnt,low}}$, $D_{\text{main}}$ & $1.137020{\times} 10^{7}$ & $1.137019{\times} 10^{7}$ & $D_{\text{pnt,low}}$, $D_{\text{outer}}$, $R_{\text{cs}}$ & $9.246560{\times} 10^{6}$ & $9.246560{\times} 10^{6}$ \\
$D_{\text{pnt,low}}$, $D_{\text{outer}}$ & $1.160375{\times} 10^{7}$ & $1.160375{\times} 10^{7}$ & $D_{\text{pnt,low}}$, $z_{\text{keel}}$, $D_{\text{main}}$ & $1.104590{\times} 10^{7}$ & $1.104587{\times} 10^{7}$  \\
$D_{\text{pnt,low}}$, $z_{\text{frbrd}}$ & $1.162365{\times} 10^{7}$& $1.162365{\times} 10^{7}$ & $D_{\text{pnt,low}}$, $D_{\text{outer}}$, $D_{\text{main}}$ & $1.119919{\times} 10^{7}$ & $1.119919{\times} 10^{7}$ \\
$D_{\text{pnt,low}}$, $R_{\text{cs}}$ & $1.163090{\times} 10^{7}$ &   $1.163090{\times} 10^{7}$ & $D_{\text{pnt,low}}$, $D_{\text{main}}$, $D_{\text{pnt,up}}$  &  $1.133210{\times} 10^{7}$ & $1.133209{\times} 10^{7}$  \\
$D_{\text{pnt,low}}$, $z_{\text{keel}}$ & $1.163325{\times} 10^{7}$ & $1.163325{\times} 10^{7}$ & $D_{\text{pnt,low}}$, $D_{\text{main}}$, $\textsf{ps}_{\text{\%}}$  &  $1.137020{\times} 10^{7}$    & $1.137019{\times} 10^{7}$  \\
$D_{\text{pnt,low}}$, $D_{\text{pnt,up}}$ & $1.163828{\times} 10^{7}$ & $1.163828{\times} 10^{7}$ &  $D_{\text{pnt,low}}$, $D_{\text{outer}}$, $z_{\text{keel}}$  & $1.137900{\times} 10^{7}$ & $1.137896{\times} 10^{7}$ \\
$D_{\text{pnt,low}}$, $\textsf{ps}_{\text{\%}}$ & $1.166550{\times} 10^{7}$ & $1.166550{\times} 10^{7}$ &  $D_{\text{pnt,low}}$, $z_{\text{keel}}$, $R_{\text{cs}}$ & $1.163026{\times} 10^{7}$  & $1.163026{\times} 10^{7}$  \\
$z_{\text{frbrd}}$, $D_{\text{main}}$ & $1.501410{\times} 10^{7}$ & $1.501409{\times} 10^{7}$& $D_{\text{pnt,low}}$, $z_{\text{frbrd}}$, $D_{\text{main}}$ & $1.108340{\times} 10^{7}$  &  $1.108339{\times} 10^{7}$ \\
$z_{\text{keel}}$, $D_{\text{main}}$  & $1.384680{\times} 10^{7}$ & $1.384680{\times} 10^{7}$ &  $\textsf{ps}_{\text{\%}}$ , $z_{\text{keel}}$, $D_{\text{main}}$ & $1.384680{\times} 10^{7}$  & $1.384680{\times} 10^{7}$ \\
$D_{\text{pnt,up}}$, $R_{\text{cs}}$   & $1.492680{\times} 10^{7}$ & $1.492680{\times} 10^{7}$  &  $\textsf{ps}_{\text{\%}}$ , $D_{\text{pnt,up}}$, $R_{\text{cs}}$ & $1.492680{\times} 10^{7}$  &  $1.492680{\times} 10^{7}$\\
$D_{\text{pnt,up}}$, $\textsf{ps}_{\text{\%}}$   & $1.509260{\times} 10^{7}$ & $1.509258{\times} 10^{7}$ &  $\textsf{ps}_{\text{\%}}$ , $D_{\text{pnt,up}}$, $D_{\text{main}}$ & $1.501410{\times} 10^{7}$  &  $1.501406{\times} 10^{7}$ \\ \midrule
$D_{\text{pnt,low}}$, $D_{\text{outer}}$, $z_{\text{keel}}$, $R_{\text{cs}}$ &  $8.006770{\times} 10^{6}$ &  $8.006771{\times} 10^{6}$  & $D_{\text{pnt,low}}$, $D_{\text{pnt,up}}$, $D_{\text{main}}$, $D_{\text{outer}}$  & $1.119410{\times} 10^{7}$ & $1.119410{\times} 10^{7}$ \\
$D_{\text{pnt,low}}$, $D_{\text{outer}}$, $D_{\text{main}}$,  $R_{\text{cs}}$ &  $8.794410{\times} 10^{6}$ &  $8.794407{\times} 10^{6}$ & $D_{\text{pnt,low}}$, $D_{\text{pnt,up}}$, $D_{\text{main}}$, $z_{\text{frbrd}}$  & $1.086450{\times} 10^{7}$ & $1.086451{\times} 10^{7}$ \\
$D_{\text{pnt,low}}$, $D_{\text{main}}$, $z_{\text{keel}}$, $R_{\text{cs}}$ & $1.072036{\times} 10^{7}$ & $1.072036{\times} 10^{7}$ &$\textsf{ps}_{\text{\%}}$ , $z_{\text{keel}}$, $z_{\text{frbrd}}$, $D_{\text{main}}$   & $1.318620{\times} 10^{7}$  & $1.318620{\times} 10^{7}$ \\
$D_{\text{pnt,low}}$, $D_{\text{outer}}$, $D_{\text{main}}$, $z_{\text{keel}}$  & $1.080170{\times} 10^{7}$ & $1.080170{\times} 10^{7}$ & $\textsf{ps}_{\text{\%}}$ , $D_{\text{pnt,up}}$, $R_{\text{cs}}$,  $z_{\text{keel}}$  & $1.329680{\times} 10^{7}$  & $1.329678{\times} 10^{7}$ \\ 
$D_{\text{pnt,low}}$, $D_{\text{outer}}$, $z_{\text{frbrd}}$, $z_{\text{keel}}$   & $1.109830{\times} 10^{7}$  & $1.109828{\times} 10^{7}$& $\textsf{ps}_{\text{\%}}$ , $D_{\text{pnt,up}}$, $D_{\text{main}}$, $z_{\text{frbrd}}$  & $1.501400{\times} 10^{7}$ & $1.501403{\times} 10^{7}$ \\
$D_{\text{pnt,low}}$, $D_{\text{pnt,up}}$, $D_{\text{main}}$, $z_{\text{keel}}$ & $1.110780{\times} 10^{7}$ & $1.110785{\times} 10^{7}$ \\
\bottomrule
\end{tabular}
}%
\end{table*}

The results for optimizing two design variables are summarized in the top left quadrant of Table~\ref{tab:weis}. The objective function values show that, although $D_{\text{main}}$ has a smaller impact on the objective function value than $z_{\text{keel}}$, optimizing $D_{\text{pnt,low}}$ together with $D_{\text{main}}$ achieves significantly lower platform mass ($1.137020 \times 10^{7}$ kg) than the combination of $D_{\text{pnt,low}}$ and $z_{\text{keel}}$ ($1.163325 \times 10^{7}$ kg). These results highlight the importance of accounting for design coupling among design variables in addition to direct impact on the objective function value.

Extending this approach to three and four design variables further reinforces this finding. The results for optimizing three design variables are summarized in the top right quadrant of Tab.~\ref{tab:weis}. Selecting the three variables with the highest impact on the objective function value, namely $D_{\text{pnt,low}}$, $z_{\text{keel}}$ and $D_{\text{outer}}$, result in a larger value in the platform mass ($1.137900 \times 10^{7}$ kg). In contrast, selecting the single variable with the highest impact, namely $D_{\text{pnt,low}}$, and the two variables that are strongly coupled both to it and to each other, namely $D_{\text{outer}}$ and $R_{\text{cs}}$, achieved the lowest value in the platform mass ($9.246560 \times 10^{6}$ kg) among all cases with three plant variable combinations. These results demonstrate that strong design coupling within a selected group of design variables can enhance optimization effectiveness beyond what direct sensitivities solely would suggest.

The results for optimizing four design variables are summarized in the bottom half (both left and right sides) of Tab.~\ref{tab:weis}. For this case, the most favorable result is achieved when optimizing $z_{\text{keel}}$, $D_{\text{outer}}$, $R_{\text{cs}}$, and $D_{\text{pnt,low}}$, resulting in the objective function value of $8.006770 \times 10^{6}$ kg. Notably, this variable combination aligns with the variable grouping selected in the second stage of Case Study 1 and yields a closely comparable objective function value, providing consistent validation of the coupling-informed strategy.

By analyzing the objective function values for different combinations of design variables, given in Table~\ref{tab:weis}, it becomes evident that, for the same number of optimized variables, the achieved performance can vary substantially depending on how the design variables are selected and grouped. In extreme cases, the difference approaches an order of magnitude. Overall, these results clearly demonstrate that effective identification of influential design variables for design optimization requires considerations of both the objective sensitivity and the design coupling matrices. The results from the four variable case in Case Study 2 closely align with the grouping identified in stage two of Case Study 1, providing consistent cross-validation of the coupling informed design variable selection framework. Relying solely on the objective sensitivities neglects indirect effects that groups of design variables have on the objective function value through design coupling.

The results highlight the value of investing in DCA to guide the selection of influential design variables and to provide general insight into design mechanisms inherent in optimization problem formulation. It should be noted that the accuracy of both the objective sensitivity and design coupling Jacobian matrices depends on the fidelity of the surrogate model. If the surrogate model does not accurately capture the underlying input to output relationships among all variables of interest, the resulting sensitivities may not reliably represent the true system behavior.

\subsection{Summary of Results and Their Computational Time}

This section compares the objective function values, along with the computational time in seconds, across all strategies in both Case Studies 1 and 2. As shown in Table~\ref{tab:grouped}, the simultaneous optimization achieves the lowest objective function value ($7.740970 \times 10^{6}$ kg) but at the highest computational cost (23,782 s). In contrast, reduced variable strategies significantly lower computational time (1,751 to 7,960 s) while still improving performance ($8.006770 \times 10^{6}$ to $1.137020 \times 10^{7}$ kg) relative to the baseline design ($1.509260 \times 10^{7}$ kg). The four variable and DCA-based sequential approaches both reach $8.006770 \times 10^{6}$ kg, with each requiring significantly less computational time than the full simultaneous optimization. Meanwhile, the random sequential approach performs considerably worse objective function value on average, highlighting the importance of coupling informed variable selection for achieving an efficient balance between accuracy and computational effort.

\begin{table}[t]
\centering
\caption{Comparison of {RAFT} optimization results and computational time for each strategy.}
\label{tab:grouped}
\begin{tabular}{@{}lrrrrr@{}}
\toprule
& Platform & Computational \\
& Mass & Time \\
Optimization Strategies & (kg) & (second) \\
\midrule
Two Variables        & $1.137020{\times} 10^{7}$  & 7,960 \\
Three Variables      & $9.246560{\times} 10^{6}$   & 1,751  \\
Four Variables       & $8.006770{\times} 10^{6}$    & 2,489  \\
DCA-based Sequential & $8.006770{\times} 10^{6}$ & 5,650  \\
Random Sequential (mean) & $1.174690{\times} 10^{7}$  & - \\
\midrule
Simultaneous         & $7.740970{\times} 10^{6}$    & 23,782 \\
Baseline Design      & $1.509260{\times} 10^{7}$  & - \\
\bottomrule
\end{tabular}
\end{table}

\section{Conclusions}
\label{sec:conclusions}

This work presents a DCA framework for FOWT optimization, with the goals of systematically understanding variable interdependencies and reformulating large multidisciplinary design problems into more tractable subproblems. The framework leverages coupling information to reveal the underlying design mechanisms that govern optimal solutions and to guide effective and tractable optimization strategies. The computational expense of evaluating coupled aero-hydro-servo-elastic FOWT models make direct calculation of design coupling and objective sensitivity Jacobian matrices impractical. To enable this analysis, a surrogate representation of the system is employed to efficiently approximate model responses and facilitate sensitivity estimations. The resulting coupling information is then used to characterize control-plant, plant-control, and plant-plant design variable interactions, and guide optimization problem formulations.

Two DCA-based optimization strategies were developed and evaluated. The first strategy, detailed in Sec.~\ref{sec:result-case-sequence}, decomposes the full simultaneous optimization problem into a sequential framework based on coupled relationships among design variables, preserving dominant design variable interdependencies while reducing problem size at each stage. The second strategy, detailed in Sec.~\ref{sec:result-case-subsetoptim}, combines objective sensitivity information with design coupling information to identify the most influential subsets of design variables for reduced-dimensional optimization. Together, these methodologies provide systematic approaches for managing large problems under limited computational resources, while maintaining critical design variable interactions and achieving optimal solutions comparable to those obtained from full simultaneous optimization.

The results demonstrate that DCA-based strategies achieve near-optimal solution at substantially lower computational expenses, if DCA information can be provided. Furthermore, variable groupings identified through coupling analysis consistently outperformed randomly selected combinations, with objective values differing by nearly an order of magnitude in extreme cases. These findings confirm that accounting for both direct sensitivities and inter-variable couplings is critical for effective optimization problem formulation.

Beyond numerical performance, this study highlights the broader value of DCA as a decision support tool for multidisciplinary design problems. Decomposition informed by design variable couplings reduces the risk of suboptimal variable selection, improves computational tractability, and provides deeper insight into the underlying design mechanisms governing FOWT systems.

In this study, the DCA information was obtained using a surrogate representation of the FOWT system. Although training the surrogate model requires a significant amount of computational resources, this effort remains substantially lower than performing the full simultaneous optimization directly on the WEIS model. Developing approaches to obtain highly reliable DCA information with reduced computational cost represents a promising direction for future research beyond the scope of this work. Furthermore, extending the DCA framework for more complex control design problems, such as optimal control problems involving time-varying or frequency-domain harmonic systems with higher-dimensional design spaces, constitutes another important avenue for future investigation.

\section*{Acknowledgment}

The authors gratefully acknowledge the use of the \textsf{BigBlue} high performance computing (HPC) facility provided by the University of Memphis (UofM) and the \textsf{Kestrel} HPC system provided by the National Laboratory of the Rockies (NLR), both of which were instrumental in supporting the computational aspects of this research.

\section*{Funding Data}

The information, data, or work presented herein was funded in part by the U.S. Department of Energy, Advanced Research Projects Agency-Energy (ARPA-E) under grant numbers DE-AR-0001182 and DE-AR-0001766.

\bibliographystyle{asmejour}
\bibliography{references}

@inproceedings{Chinthoju2024eVTOL,
    author = {Chinthoju, Prajwal K. and Lee, Yong Hoon and Das, Ghanendra K. and James, Kai A. and Allison, James T.},
    title = {Optimal Design of {eVTOLs} for Urban Mobility Using Analytical Target Cascading ({ATC})},
    booktitle = {Proceedings of the AIAA SciTech Forum and Exposition},
    year = {2024},
    eventdate = {January 8--12},
    venue = {Orlando, FL, USA},
    number = {AIAA 2024-2235},
    pages = {1--13},
    doi = {10.2514/6.2024-2235},
}

@inproceedings{FernandezBravo2024SatelliteMDO,
    author = {Fern{\'a}ndez Bravo, Elena and Allison, James T.},
    title = {Satellite {MDO} Problem Formulation Using Design Coupling Information},
    booktitle = {Proceedings of the 75th International Astronautical Congress},
    year = {2024},
    eventdate = {October 14--18},
    venue = {Milan, Italy},
    number = {88072},
    pages = {},
    url = {https://iafastro.directory/iac/archive/browse/IAC-24/D1/IPB/88072/},
}

@article{Pao2024AnnuRev,
    author = {Pao, Lucy Y. and Pusch, Manuel and Zalkind, Daniel S.},
    title = {Control Co-Design of Wind Turbines}, 
    journal= {Annual Review of Control, Robotics, and Autonomous Systems},
    year = {2024},
    volume = {7},
    number = {--},
    pages = {201--226},
    doi = {10.1146/annurev-control-061423-101708},
}

@article{Lee2025FloatVAWT,
    author = {Lee, Yong Hoon and Bayat, Saeid and Allison, James T. and Hossain, Md Sanower and Griffith, D. Todd},
    title = {Multidisciplinary Modeling and Control Co-Design of a Floating Offshore Vertical-Axis Wind Turbine System},
    journal = {Journal of Mechanical Design},
    year = {2025},
    volume = {147},
    number = {6},
    pages = {061702},
    doi = {10.1115/1.4068072},
}

@article{Lee2025AppliedEnergy,
    author = {Lee, Yong Hoon and Bayat, Saeid and Allison, James T.},
    title = {Wind Turbine Control Co-Design Using Dynamic System Derivative Function Surrogate Model ({DFSM}) Based on {OpenFAST} Linearization},
    journal = {Applied Energy},
    year = {2025},
    volume = {396},
    number = {--},
    pages = {126203},
    doi = {10.1016/j.apenergy.2025.126203},
}

@article{Bayat2025OE,
    author = {Bayat, Saeid and Lee, Yong Hoon and Allison, James T.},
    title = {Nested Control Co-Design of a Spar Buoy Horizontal-Axis Floating Offshore Wind Turbine},
    journal = {Ocean Engineering},
    year = {2025},
    volume = {328},
    number = {--},
    pages = {121037},
    doi = {10.1016/j.oceaneng.2025.121037},
}

@article{sundarrajanOpenLoopControlCoDesign2024,
	author = {Sundarrajan, Athul K. and Lee, Yong Hoon and Allison, James T. and Zalkind, Daniel S. and Herber, Daniel R.},
    title = {Open-Loop Control Co-Design of Semisubmersible Floating Offshore Wind Turbines Using Linear Parameter-Varying Models},
    journal = {ASME Journal of Mechanical Design},
	year = {2024},
	volume = {146},
	number = {4},
    pages = {041704},
	doi = {10.1115/1.4063969},
}

@article{ABBAS2024ccd,
    author = {Abbas, Nikhar J. and Jasa, John and Zalkind, Daniel S. and Wright, Alan and Pao, Lucy},
    title = {Control Co-Design of a Floating Offshore Wind Turbine},
    journal = {Applied Energy},
    year = {2024},
    volume = {353},
    number = {--},
    pages = {122036},
    doi = {10.1016/j.apenergy.2023.122036},
}

@article{Cui2021,
    author = {Cui, Tonghui and Allison, James T. and Wang, Pingfeng},
    title = {Reliability-Based Control Co-Design of Horizontal Axis Wind Turbines},
    journal = {Structural and Multidisciplinary Optimization},
    year = {2021},
    volume = {64},
    number = {6},
    pages = {3653--3679},
    doi = {10.1007/s00158-021-03046-3},  
}

@article{Sakib2024,
    author = {Sakib, Mohammad Sadman and Griffith, D. Todd and Hossain, Sanower and Bayat, Saeid and Allison, James T.},
    title = {Intracycle {RPM} Control for Vertical Axis Wind Turbines},
    journal = {Wind Energy},
    year = {2024},
    volume = {27},
    number = {3},
    pages = {202--224},
    doi = {10.1002/we.2885},
}

@article{Deshmukh2016,
    author = {Deshmukh, Anand P. and Allison, James T.},
    title = {Multidisciplinary Dynamic Optimization of Horizontal Axis Wind Turbine Design},
    journal = {Structural and Multidisciplinary Optimization},
    year = {2016},
    volume = {53},
    number = {1},
    pages = {15--27},
    doi = {10.1007/s00158-015-1308-y},
}

@article{MengKrigingAssistedRBMDO2023,
    author = {Meng, Debiao and Yang, Shiyuan and de Jesus, Abílio M. P. and Zhu, Shun-Peng},
    title = {A Novel Kriging-Model-Assisted Reliability-Based Multidisciplinary Design Optimization Strategy and Its Application in the Offshore Wind Turbine Tower},
    journal = {Renewable Energy},
    year = {2023},
    volume = {203},
    number = {--},
    pages = {407--420},
    doi = {10.1016/j.renene.2022.12.062},
}

@article{ChenReviewOffshoreWind2022,
    author = {Chen, Jieyan and Kim, Moo-Hyun},
    title = {Review of Recent Offshore Wind Turbine Research and Optimization Methodologies in Their Design},
    journal = {Journal of Marine Science and Engineering},
    year = {2022},
    volume = {10},
    number = {1},
    pages = {28},
    doi = {10.3390/jmse10010028},
}

@article{ChoeSequenceBasedModeling2021,
    author = {Choe, Do-Eun and Kim, Hyoung-Chul and Kim, Moo-Hyun},
    title = {Sequence-Based Modeling of Deep Learning with {LSTM} and {GRU} Networks for Structural Damage Detection of Floating Offshore Wind Turbine Blades},
    journal = {Renewable Energy},
    year = {2021},
    volume = {174},
    number = {--},
    pages = {218--235},
    doi = {10.1016/j.renene.2021.04.025},
}

@article{FengDesignOptimizationOffshore2017,
    author = {Feng, Ju and Shen, Wen Zhong},
    title = {Design Optimization of Offshore Wind Farms with Multiple Types of Wind Turbines},
    journal = {Applied Energy},
    year = {2017},
    volume = {205},
    number = {--},
    pages = {1283--1297},
    doi = {10.1016/j.apenergy.2017.08.107},
}

@article{MuskulusDesignOptimizationSupport2014,
    author = {Muskulus, Michael and Schafhirt, Sebastian},
    title = {Design Optimization of Wind Turbine Support Structures: A Review},
    journal = {Journal of Ocean and Wind Energy},
    year = {2014},
    volume = {1},
    number = {1},
    pages = {12--22},
    doi = {},
    internalcomment = {DOI number does not exist.},
}

@inproceedings{fathyCouplingPlantController2001,
	author = {Fathy, H. K. and Reyer, J. A. and Papalambros, P. Y. and Ulsov, A. G.},
	title = {On the Coupling Between the Plant and Controller Optimization Problems},
	booktitle = {Proceedings of the 2001 American Control Conference},
	year = {2001},
    eventdate = {June 25--27},
	venue = {Arlington, VA, USA},
    volume = {3},
	pages = {1864--1869},
	doi = {10.1109/ACC.2001.946008},
}

@phdthesis{fathyCombinedPlantControl2003,
	author = {Fathy, Hosam K},
	title = {Combined Plant and Control Optimization: Theory, Strategies and Applications},
	school = {University of Michigan},
	year = {2003},
	type = {Ph.{D}. dissertation},
    address = {Ann Arbor, MI, USA},
}

@article{shabdeOptimumControllerDesign2008,
	author = {Shabde, Vikram S. and Hoo, Karlene A.},
	title = {Optimum Controller Design for a Spray Drying Process},
	journal = {Control Engineering Practice},
	year = {2008},
	volume = {16},
	number = {5},
	pages = {541--552},
	doi = {10.1016/j.conengprac.2007.06.004},
}

@article{ulsoySmartProductDesign2019,
	author = {Ulsoy, A. Galip},
	title = {Smart Product Design for Automotive Systems},
	journal = {Frontiers of Mechanical Engineering},
	year = {2019},
	volume = {14},
	number = {1},
	pages = {102--112},
	doi = {10.1007/s11465-019-0527-0},
}

@inproceedings{fathyCombinedPlantControl2004,
	author = {Fathy, H. K. and Papalambros, P. Y. and Ulsoy, A.G.},
	title = {On Combined Plant and Control Optimization},
	booktitle = {Proceedings of the 8th Cairo University International Conference on Mechanical Design and Production},
	year = {2004},
    eventdate = {January 4--6},
	venue = {Cairo, Egypt},
    number = {},
    pages = {1--9},
    doi = {},
}

@article{petersRelationshipCouplingControllability2015,
	author = {Peters, Diane L. and Papalambros, Panos Y. and Ulsoy, A. Galip},
	title = {Relationship {Between} {Coupling} and the {Controllability} {Grammian} in {Co}-{Design} {Problems}},
	journal = {Mechatronics},
	year = {2015},
	volume = {29},
    number = {--},
    pages = {36--45},
	doi = {10.1016/j.mechatronics.2015.05.002},
}

@inproceedings{petersMeasuresCouplingArtifact2009,
	author = {Peters, Diane L. and Papalambros, Panos Y. and Ulsoy, A. Galip},
	title = {On Measures of Coupling Between the Artifact and Controller Optimal Design Problems},
	booktitle = {Proceedings of the ASME 2009 International Design Engineering Technical Conferences and Computers and Information in Engineering Conference},
	year = {2009},
    eventdate = {August 30--September 2},
    venue = {San Diego, CA, USA},
	number = {DETC2009-86868},
	pages = {1363--1372},
    doi = {10.1115/DETC2009-86868},
}

@inproceedings{fernandezbravoNumericalEstimationBidirectional2024,
	author = {Fern{\'a}ndez Bravo, Elena and Ornik, Melkior and Allison, James T.},
	title = {Numerical Estimation of Bidirectional Plant-Control Design Coupling in Control Co-Design},
	booktitle = {Proceedings of the ASME 2024 International Design Engineering Technical Conferences and Computers and Information in Engineering Conference},
	year = {2024},
	eventdate = {August 25--28},
	venue = {Washington, DC, USA},
    number = {DETC2024-142636},
    pages = {V03AT03A002},
	doi = {10.1115/DETC2024-142636},
}

@article{Wang2020SBCCD,
    author = {Wang, Xiaobang and Song, Xueguan and Sun, Wei and Sun, Changguo and Liu, Zhijie},
    title = {Surrogate Based Co-Design for Combined Structure and Control Design Problems}, 
    journal = {IEEE Access}, 
    year = {2020},
    volume = {8},
    number = {--},
    pages = {184851--184865},
    doi = {10.1109/ACCESS.2020.3029390},
}

@article{qiaoNewSequentialSampling2021,
	author = {Qiao, Ping and Wu, Yizhong and Ding, Jianwan and Zhang, Qi},
	title = {A new sequential sampling method of surrogate models for design and optimization of dynamic systems},
	journal = {Mechanism and Machine Theory},
	year = {2021},
	volume = {158},
    number = {--},
    pages = {104248},
	doi = {10.1016/j.mechmachtheory.2021.104248},
}

@article{zhangNovelSurrogateModelBased2022,
	author = {Zhang, Qi and Wu, Yizhong and Lu, Li},
	title = {A Novel Surrogate Model-Based Solving Framework for the Black-Box Dynamic Co-Design and Optimization Problem in the Dynamic System},
	journal = {Mathematics},
	year = {2022},
	volume = {10},
	number = {18},
	pages = {3239},
	doi = {10.3390/math10183239},
}

@inproceedings{sundarrajanUsingHighFidelityTimeDomain2023,
	author = {Sundarrajan, Athul and Herber, Daniel R.},
	title = {Using {High}-{Fidelity} {Time}-{Domain} {Simulation} {Data} to {Construct} {Multi}-{Fidelity} {State} {Derivative} {Function} {Surrogate} {Models} for {Use} in {Control} and {Optimization}},
    booktitle = {Proceedings of the ASME 2023 International Mechanical Engineering Congress and Exposition},
	year = {2023},
    eventdate = {October 29–November 2},
	venue = {New Orleans, LA, USA},
    number = {IMECE2023-112316},
    pages = {V006T07A089},
    doi = {10.1115/IMECE2023-112316},
}

@article{Deshmukh2017DFSM,
    author = {Deshmukh, Anand P. and Allison, James T.},
	title = {Design of Dynamic Systems Using Surrogate Models of Derivative Functions},
    journal = {Journal of Mechanical Design},
    year = {2017},
    volume = {139},
    number = {10},
    pages = {101402},
    doi = {10.1115/1.4037407},
}

@article{Lee2019SMO,
    author = {Lee, Yong Hoon and Schuh, Jonathon K. and Ewoldt, Randy H. and Allison, James T.},
    title = {Simultaneous Design of Non-{N}ewtonian Lubricant and Surface Texture Using Surrogate-Based Multiobjective Optimization},
    journal = {Structural and Multidisciplinary Optimization},
    year = {2019},
    volume = {60},
    number = {1},
    pages = {99--116},
    doi = {10.1007/s00158-019-02201-1},
}

@article{Malak2010SVDD,
    author = {Malak, Richard J., Jr. and Paredis, Christiaan J. J.},
    title = {Using Support Vector Machines to Formalize the Valid Input Domain of Predictive Models in Systems Design Problems},
    journal = {Journal of Mechanical Design},
    year = {2010},
    volume = {132},
    number = {10},
    pages = {101001},
    doi = {10.1115/1.4002151},
}

@article{LucasFrutuoso2025AdvWindFarm,
    author = {Lucas Frutuoso, Tiago R. and Castro, Rui and Pereira, Ricardo B. Santos and Moutinho, Alexandra},
    title = {Advancements in Wind Farm Control: Modelling and Multi-Objective Optimization Through Yaw-Based Wake Steering},
    journal = {Energies},
    year = {2025},
    volume = {18},
    number = {9},
    pages = {2247},
    doi = {10.3390/en18092247},
}

@inproceedings{jonkmanFunctionalRequirementsWEIS2021,
	authorfull = {Jonkman, Jason and Wright, Alan and Barter, Garrett and Hall, Matthew and Allison, James and Herber, Daniel R.},
    author = {Jonkman, Jason and Wright, Alan and Barter, Garrett and others},
	title = {Functional Requirements for the {WEIS} Toolset to Enable Controls Co-Design of Floating Offshore Wind Turbines},
	booktitle = {Proceedings of the ASME 2021 3rd International Offshore Wind Technical Conference},
	year = {2021},
    eventdate = {February 16--17},
	venue = {Virtual, Online},
    number = {IOWTC2021-3533},
    pages = {V001T01A007},
	doi = {10.1115/IOWTC2021-3533},
}

@incollection{Matha2016FOWT,
    authorfull = {Matha, Denis and Cruz, Joao and Masciola, Marco and Bachynski, Erin E. and Atcheson, Mair{\'e}ad and Goupee, Andrew J. and Gueydon, S{\'e}bastien M. H. and Robertson, Amy N.},
    author = {Matha, Denis and Cruz, Joao and Masciola, Marco and others},
    title = {Modelling of Floating Offshore Wind Technologies},
    booktitle = {Floating Offshore Wind Energy: The Next Generation of Wind Energy},
    editor = {Cruz, Joao and Atcheson, Mairead},
    year = {2016},
    publisher = {Springer},
    address = {Cham},
    pages = {133--240},
    doi = {10.1007/978-3-319-29398-1_4},
}

@article{Johlas2019LES,
    author = {Johlas, H. M. and Mart{\'i}nez-Tossas, L. A. and Schmidt, D. P. and Lackner, M. A. and Churchfield, M. J.},
    title = {Large Eddy Simulations of Floating Offshore Wind Turbine Wakes With Coupled Platform Motion},
    journal = {Journal of Physics: Conference Series},
    year = {2019},
    volume = {1256},
    number = {1},
    pages = {012018},
    doi = {10.1088/1742-6596/1256/1/012018},
}

@article{CampanaAlonso2023OF2,
    author = {Campa{\~n}a-Alonso, Guill{\'e}n and Mart{\'i}n-San-Rom{\'a}n, Raquel and M{\'e}ndez-L{\'o}pez, Beatriz and Benito-Cia, Pablo and Azcona-Armend{\'a}riz, Jos{\'e}},
    title = {{OF}$^2$: Coupling {OpenFAST} and {OpenFOAM} for High-Fidelity Aero-Hydro-Servo-Elastic {FOWT} Simulations},
    journal = {Wind Energy Science},
    year = {2023},
    volume = {8},
    number = {10},
    pages = {1597--1611},
    doi = {10.5194/wes-8-1597-2023},
}

@inproceedings{Jonkman2013FAST,
    author = {Jonkman, Jason},
    title = {The New Modularization Framework for the {FAST} Wind Turbine {CAE} Tool},
    booktitle = {51st AIAA Aerospace Sciences Meeting including the New Horizons Forum and Aerospace Exposition},
    year = {2013},
    eventdate = {January 7--10},
    venue = {Grapevine, TX, USA},
    number = {AIAA 2013-0202},
    pages = {1--26},
    doi = {10.2514/6.2013-202},
}

@phdthesis{Marten2020QBlade,
    author = {Marten, David},
    title = {{QBlade}: A Modern Tool for the Aeroelastic Simulation of Wind Turbines},
    school = {Technischen Universit{\:a}t Berlin},
    year = {2020},
    type = {Ph.{D}. dissertation},
    address = {Berlin, Germany},
    doi = {10.14279/depositonce-10646},
}

@article{Beardsell2018Bladed,
    author = {Beardsell, A. and Alexandre, A. and Child, B. and Harries, R. and McCowen, D.},
    title = {Beyond {OC5} -- Further Advances in Floating Wind Turbine Modelling Using {Bladed}},
    journal = {Journal of Physics: Conference Series},
    year = {2018},
    volume = {1102},
    number = {1},
    pages = {012023},
    doi = {10.1088/1742-6596/1102/1/012023},
}

@inproceedings{JonkmanFullSystemOpenFAST2018,
    author    = {Jonkman, Jason M. and Wright, Alan D. and Hayman, Greg J. and Robertson, Amy N.},
    title     = {Full-System Linearization for Floating Offshore Wind Turbines in {OpenFAST}},
    booktitle = {Proceedings of the ASME 2018 1st International Offshore Wind Technical Conference},
    year      = {2018},
    eventdate = {November 4--7},
    venue = {San Francisco, CA, USA},
    number = {IOWTC2018-1025},
    pages = {V001T01A028},
    doi = {10.1115/IOWTC2018-1025},
}

@article{hallOpenSourceFrequencyDomainModel2022,
    authorfull = {Hall, Matthew and Housner, Stein and Zalkind, Daniel and Bortolotti, Pietro and Ogden, David and Barter, Garrett},
    author = {Hall, Matthew and Housner, Stein and Zalkind, Daniel and others},
	title = {An Open-Source Frequency-Domain Model for Floating Wind Turbine Design Optimization},
    journal = {Journal of Physics: Conference Series},
	year = {2022},
	volume = {2265},
    number = {4},
    pages = {042020},
	doi = {10.1088/1742-6596/2265/4/042020},
}

@article{OtterReviewFOWT2021,
    author = {Otter, Aldert and Murphy, Jimmy and Pakrashi, Vikram and Robertson, Amy and Desmond, Cian},
    title = {A Review of Modelling Techniques for Floating Offshore Wind Turbines},
    journal = {Wind Energy},
    year = {2021},
    volume = {25},
    number = {5},
    pages = {963--983},
    doi = {10.1002/we.2701},
}

@techreport{Gaertner2020IEA15MWrefturb,
    authorfull = {Gaertner, Evan and Rinker, Jennifer and Sethuraman, Latha and Zahle, Frederik and Anderson, Benjamin and Barter, Garrett and Abbas, Nikhar and Meng, Fanzhong and Bortolotti, Pietro and Skrzypinski, Witold and Scott, George and Feil, Roland and Bredmose, Henrik and Dykes, Katherine and Shields, Matt and Allen, Christopher and Viselli, Anthony},
    author = {Gaertner, Evan and Rinker, Jennifer and Sethuraman, Latha and others},
    title = {Definition of the {IEA} Wind 15-Megawatt Offshore Reference Wind Turbine},
    institution = {National Renewable Energy Laboratory},
    type = {Technical Report},
    year = {2020},
    month = mar,
    number = {NREL/TP-5000-75698},
    address = {Golden, CO, USA},
    url = {https://docs.nrel.gov/docs/fy20osti/75698.pdf},
}

@techreport{Allen2020UMaineSemiRef,
    authorfull = {Allen, Christopher and Viselli, Anthony and Dagher, Habib and Goupee, Andrew and Gaertner, Evan and Abbas, Nikhar and Hall, Matthew and Barter, Garrett},
    author = {Allen, Christopher and Viselli, Anthony and Dagher, Habib and others},
    title = {Definition of the {UMaine} {VolturnUS-S} Reference Platform Developed for the {IEA} Wind 15-Megawatt Offshore Reference Wind Turbine},
    institution = {National Renewable Energy Laboratory},
    type = {Technical Report},
    year = {2020},
    month = jul,
    number = {NREL/TP-5000-76773},
    address = {Golden, CO, USA},
    url = {https://www.nrel.gov/docs/fy20osti/76773.pdf},
}

@article{Liu2019HAMS,
    author = {Liu, Yingyi},
    title = {{HAMS}: A Frequency-Domain Preprocessor for Wave-Structure Interactions—Theory, Development, and Application},
    journal = {Journal of Marine Science and Engineering},
    year = {2019},
    volume = {7},
    number = {3},
    pages = {81},
    doi = {10.3390/jmse7030081},
}

@phdthesis{herberAdvancesCombinedArchitecture,
	author = {Herber, Daniel Ronald},
	title = {Advances in {Combined} {Architecture}, {Plant}, and {Control} {Design}},
    school = {University of Illinois at Urbana-Champaign},
    year = {2017},
    type = {Ph.{D}. dissertation},
    address = {Urbana, IL, USA},
    url = {https://hdl.handle.net/2142/99394},
}

@misc{Herber2017DTQPproject,
    author = {Herber, Daniel R. and Lee, Yong Hoon and Allison, James T.},
    title = {DT QP Project},
    year = {2017},
    howpublished = {[Computer Software] \url{https://github.com/danielrherber/dt-qp-project}},
}

@article{Jasa2022WEISmultifidelity,
    author = {Jasa, J. and Bortolotti, P. and Zalkind, D. and Barter, G.},
    title = {Effectively Using Multifidelity Optimization for Wind Turbine Design},
    journal = {Wind Energy Science},
    year = {2022},
    volume = {7},
    number = {3},
    pages = {991--1006},
    doi = {10.5194/wes-7-991-2022},
}

@article{hallFrequencyDomainModelingFloating2025,
	author = {Hall, Matthew and Carmo, Lucas and Lozon, Ericka},
	title = {Frequency-Domain Modeling of Floating Wind Arrays with Shared Mooring Lines},
    journal = {Wind Energy Science},
	year = {2025},
    volume = {10},
    number = {12},
    pages = {3027--3043},
	doi = {10.5194/wes-10-3027-2025},
}

@article{morisonForceExertedSurface1950,
	author = {Morison, J. R. and Johnson, J. W. and Schaaf, S. A.},
	title = {The Force Exerted by Surface Waves on Piles},
	journal = {Journal of Petroleum Technology},
	year = {1950},
	volume = {2},
	number = {05},
	pages = {149--154},
	doi = {10.2118/950149-G},
}

@misc{hallMoorPyQuasiStaticMooring2021,
	author = {Hall, Mathew and Housner, Stein and Sirnivas, Senu and Wilson, Samuel},
	title = {{MoorPy} (Quasi-Static Mooring Analysis in Python)},
	year = {2021},
    howpublished = {[Computer Software] \url{https://doi.org/10.11578/dc.20210726.1}},
    doi = {10.11578/dc.20210726.1},
}

@article{ningSimpleSolutionMethod2014,
	author = {Ning, S. Andrew},
	title = {A Simple Solution Method for the Blade Element Momentum Equations With Guaranteed Convergence},
	journal = {Wind Energy},
	year = {2014},
	volume = {17},
	number = {9},
    pages = {1327--1345},
    doi = {10.1002/we.1636},
}

@inproceedings{bauerUnderstandingProbabilisticSparse2016,
    author = {Bauer, Matthias and van der Wilk, Mark and Rasmussen, Carl Edward},
    title = {Understanding Probabilistic Sparse Gaussian Process Approximations},
    booktitle = {Advances in Neural Information Processing Systems: 30th Annual Conference on Neural Information Processing Systems 2016},
    editor = {D. Lee and others},
    year = {2016},
    eventdate = {December 5--10},
    venue = {Barcelona, Spain},
    volume = {29},
    pages = {1533--1541},
    publisher = {Curran Associates, Inc.},
    isbn = {9781510838819}, 
}

@article{savesSMT20Surrogate2024,
	authorfull = {Saves, Paul and Lafage, R{\'e}mi and Bartoli, Nathalie and Diouane, Youssef and Bussemaker, Jasper and Lefebvre, Thierry and Hwang, John T. and Morlier, Joseph and Martins, Joaquim R. R. A.},
    author = {Saves, Paul and Lafage, R{\'e}mi and Bartoli, Nathalie and others},
	title = {{SMT} 2.0: A Surrogate Modeling Toolbox With a Focus on Hierarchical and Mixed Variables {G}aussian Processes},
	journal = {Advances in Engineering Software},
	year = {2024},
	volume = {188},
    number = {--},
	pages = {103571},
	doi = {10.1016/j.advengsoft.2023.103571},
}

@article{Bayat2026wtccdapen,
    author = {Bayat, Saeid and Peterson, Chad and Lee, Yong Hoon and Iori, Jenna and Allison, James T.},
    title = {Advancing wind turbines through control co-design: An integrative review},
    journal = {Applied Energy},
    year ={2026},
    volume = {},
    number = {},
    pages = {},
    doi = {},
    note = {in press},
}

\end{document}